\documentclass[12pt,preprint2]{aastex}
\usepackage{amsmath}

\begin{document}

\title{On the Effect of Magnetic Spots on Stellar Winds and Angular Momentum Loss}


\author{O. Cohen\altaffilmark{1}, J.J. Drake\altaffilmark{1}, V.L. Kashyap\altaffilmark{1}, 
and T.I. Gombosi\altaffilmark{2}}

\altaffiltext{1}{Harvard-Smithsonian Center for Astrophysics, 60 Garden St. Cambridge, MA 02138}
\altaffiltext{2}{Center for Space Environment Modeling, University of Michigan, 2455 Hayward St., 
Ann Arbor, MI 48109}

\begin{abstract}

We simulate the effect of latitudinal variations in the location of star spots, as well as their magnetic field 
strength, on stellar angular momentum loss to the stellar wind. We use the Michigan solar corona 
global MagnetoHydroDynamic model, which incorporates realistic relation between the magnetic 
field topology and the wind distribution. We find that the spots' location significantly affects the 
stellar wind structure, and as a result, the total mass loss rate and angular momentum loss rate. 
In particular, we find that the angular momentum loss rate is controlled by the mass flux when spots 
are located at low latitudes but is controlled by an increased plasma density between the stellar 
surface and the Alfv\'en surface when spots are located at high latitudes. Our results suggest that 
there might be a feedback mechanism between the magnetic field distribution, wind distribution, 
angular momentum loss through the wind, and the motions at the convection zone that generate the 
magnetic field. This feedback might explain the role of coronal magnetic fields in stellar dynamos.   

\end{abstract}



\section{INTRODUCTION}
\label{sec:Intro}

Angular Momentum Loss (AML) through a magnetized wind can play a significant role in stellar 
evolution. It has been used to explain a number of stellar phenomena such as the evolution of close 
binaries and cataclysmic variables (CVs), evolution of stellar rotation rates, and dynamo 
saturation \citep[e.g.][and references therein]{ivanovataam03}.

The principle mechanism by which the wind carries out angular momentum from the rotating object is as follows. 
The magnetic field is tangled to the surface of the star at one end but its other end is opened by the frozen-in 
wind at the point where the flow becomes super-Alfv\'enic (hereafter referred to in three-dimensions as the 'Alfv\'en surface'). 
A braking torque is applied because there is corotation out to some effective radius larger than the stellar radius and the 
net result is stellar spin-down \citep{parker58}. 

Another factor affecting the AML in this process is the effect of the magnetic field on the stellar wind. In the 
solar case, the magnetic field determines not only the large scale distribution of the wind (blowing 
along open field lines), but it also determines its bi-modal structure 
\citep{phillips95,lamerscassinelli99,suzuki06,wangsheeley06,cranmer07}. 
We can then infer that the distribution of stellar winds in stars with hot coronae like 
that of the Sun will be tightly 
related to the coronal magnetic field configuration . 
As a result, the mass loss and AML will also be influenced by the magnetic field 
structure and its surface distribution. 

Doppler Imaging and Zeeman Doppler Imaging (ZDI) observations of a large number of stars have revealed 
that stellar magnetic active regions, appear as dark spots ('spots' hereafter), can appear in all latitudes, including the polar 
regions \citep{strassmeier96,strassmeier01,donaticameron97}. There is also some observational evidence that the polar
regions of active stars harbor bright coronal structures \citep[e.g.][]{chung04}.
In order to understand AML in stars is is then important to study the relation between 
the distribution of stellar magnetic fields and their AML. 

\cite{weberdavis67} (WD hereafter) have shown that for an idealized, 
spherically symmetric configuration, the loss of angular momentum by the wind can be written as: 
\begin{equation}
\label{WD}
\dot{J}=\frac{3}{2}\Omega \dot{M}r_A^2 .
\end{equation}
where $\Omega$ is the rotation rate, $\dot{M}$ is the mass loss rate, and $r_A$ is the radius of 
the Alfv\'en surface. Based on WD's model, a large number of more detailed and complex studies of AML in cool and hot stars, 
as well as in protostellar accretion disks, have been done both analytically \citep{mestel68,
holzwarth05} and numerically \citep{keppensgoedbloed00,matt08a,matt08b,uddoula09}. In particular, 
\cite{aibeo07} have shown that AML can be modified by a factor of 2-4 when introducing magnetic 
flux at high latitudes. They explain the change in AML as a result of a change in the mass flux or the size of the 
Alfv\'en surface, but they argue that the change in the magnetic flux tubes topology should not 
significantly affect magnetic breaking.

In all of the studies mentioned above, the wind distribution has been specified by some ad-hoc 
assumptions, though with a strong physical basis, such as Parker's wind solution. 
Such  a ``pre-defined'' 
wind can be appropriate in the case of an idealized and simplified magnetic field topology, in which the 
steady-state wind structure can be predicted. Even the well known Parker solution, which provides a 
very good first order approximation for the solar wind, is hydrodynamic in nature and assumes an idealized, 
spherically symmetric field configuration, with the flow freely moving along open field lines. 
These studies are very useful for parametric study of AML in stars, but they are similar to the 
potential field approximation \citep{altschulernewkirk69,mcIvor03,donati08}, in which strong assumptions 
about the boundary conditions have been made.

In order to compute AML in stars more realistically, one should calculate the true, 
non-spherical, 
Alfv\'en surface, as well the actual wind distribution that results 
when realistic magnetic fields with higher degree 
of complexity (i.e., with star spots) are introduced. In this work, we present an implementation of a global 
MagnetoHydroDynamic (MHD) model, originally developed for the solar corona, to the corona of a Sun-like star. This model 
is driven by 'realistic' surface magnetograms and it captures the detailed physics of the corona. 
Unlike other models, the wind in this model is determined by the magnetic field configuration in a self-consistent 
manner.  We exploit this approach as a tool to investigate the effect of the magnetic field topology on the wind 
structure and AML in a more consistent and realistic way. In particular, we focus on the effect of 
variations in star spot location and magnetic field magnitude on the wind structure, mass loss, and AML rates.

The topic of this paper covers overlapping issues from both solar and stellar astrophysics, 
in which different terminology is sometime used to describe similar processes. readers unfamiliar 
with numerical models for the solar corona should consult 
\citet{cohen07a,cohen08} for a more detailed description of the terminology employed here.

We describe the numerical model in Section~\ref{sec:Model} and present numerical tests and results in 
Section~\ref{sec:Simulation}.  In Section~\ref{sec:Discussion} we discuss the implications of the results 
over the long-term in the context of stellar evolution,
as well as for the short-term variations in the stellar magnetic field (the stellar cycle). 
We summarize our conclusions in Section~\ref{sec:conclusion}.


\section{MODEL DESCRIPTION}
\label{sec:Model}

In order to simulate the corona of a Sun-like star, we use a global MHD model 
developed for the 
Solar Corona (SC) at the University of Michigan.  We describe the model here briefly, 
and refer the reader to \cite{cohen07a,cohen08} for a more complete description, supporting observational 
data, referencing, and model validation. The model is based on the BATS-R-US generic MHD code 
\citep{powell99} and it is part of the Space Weather Modeling Framework (SWMF) \citep{toth05}.

The model solves the set of MHD equations:
\begin{align*}
\frac{\partial \rho}{\partial t}+\nabla\cdot(\rho \mathbf{u})=0,
\end{align*}
\begin{align}
\label{MHD}
\rho \frac{\partial \mathbf{u}}{\partial t}+
\nabla\cdot\left( 
\rho\mathbf{u}\mathbf{u}+p+\frac{B^2}{2\mu_0}-\frac{\mathbf{B}\mathbf{B}}{\mu_0}
\right) = \rho\mathbf{g},
\end{align}
\begin{align*}
\frac{\partial \mathbf{B}}{\partial t}+
\nabla\cdot(\mathbf{u}\mathbf{B}-\mathbf{B}\mathbf{u})=0,
\end{align*}
\begin{align*}
&\frac{\partial }{\partial t}\left( 
\frac{1}{2}\rho u^2+\frac{1}{\gamma-1}p+\frac{B^2}{2\mu_0}
 \right)+& \\
&\nabla\cdot\left( 
\frac{1}{2}\rho u^2\mathbf{u}+\frac{\gamma}{\gamma-1}p\mathbf{u}+
\frac{(\mathbf{B}\cdot\mathbf{B})-\mathbf{B}(\mathbf{B}\cdot\mathbf{u})}{\mu_0} 
\right)=& \\
&\rho(\mathbf{g}\cdot\mathbf{u}),& 
\end{align*}
which describe the conservation of mass, momentum, magnetic flux, and total energy. The equations are 
solved in a Cartesian grid with Adaptive Mesh Refinement (AMR) capabilities, while the 
condition $\nabla\cdot\mathbf{B}=0$ is maintained using the Eight Wave scheme \citep{powell99}. 
This SC model has been widely used to simulate the steady state solar corona, 
as well as Coronal Mass Ejections (CMEs) \cite[e.g.][]{cohen08,manchester08}.

\subsection{Wind Powering in the Numerical Model} 
\label{WindPowering}

In order to mimic the stellar wind powering, we adopt a non-polytropic approach where 
$\gamma=C_p/C_v$ is not constant. Here $C_p$ and $C_v$ are the specific heats of the gas. 
$\gamma$, is observed to be close to unity at the base of the solar corona, but it 
approaches the value $5/3$ above $15-20R_\odot$ \citep{totten95}. We can assume that the 
energy necessary to power the wind corresponds to this change in $\gamma$. 
We can also assume that this change in $\gamma$ is larger for the fast wind 
and smaller for the slow wind. The end result is that with a proper choice of the three-dimensional 
distribution of $\gamma$, we can mimic the volumetric heating of the wind and 
obtain a steady state MHD solution for the corona without introducing any heating 
functions.

In the solar case, we specify the distribution of $\gamma$ in a way that is consistent with 
observations.  By then using the empirical Wang-Sheeley-Arge (WSA) model \citep{wangy90,argepizzo00}, we also obtain a wind spatial velocity distribution that is consistent with the magnetic field distribution.  
The WSA model relates the magnetic flux tube expansion rate, $f_s$, to the final 
wind speed blowing along this tube. The higher the expansion is, the slower the wind that 
originates from the tube and vice versa. By using the WSA model as an input, we constrain our model to 
approach the empirical value of the wind speed.  

The WSA model has been developed to predict the solar wind speed at 1AU based on the potential 
field approximation and magnetogram data. The model converts the surface magnetic field data 
into a spherical distribution of the terminal solar wind speed by matching its parameters with 
solar wind data at 1~AU. We assume that the wind in Sun-like stars is not significantly different in its underlying physics from 
the solar case, in particular the relation between the expansion rate and the wind speed. 
In principle, the dependency of the wind speed on the 
expansion rate, $f_s$, could be fit to stellar wind observations such as those presented by \cite{wood05}.

We calculate the potential field and WSA coefficients in order to obtain the spherical 
distribution of the final wind speed. We assume conservation of energy along a streamline 
(Bernoulli Integral) so we can relate the value of the speed at one end of the field line 
far from the Sun to the value of $\gamma$ at the coronal base at the other end. If we assume 
that far from the Sun the total energy roughly equals the kinetic energy we can obtain the 
following equation:
\begin{equation}
\label{Bernoulli2}
\frac{u_{wsa}^2}{2}=\frac{\gamma_0 }{[\gamma_0-1]}
\frac{p_0}{\rho_0}-\frac{GM_\odot}{R_\odot},
\end{equation}
where G is the gravitational constant, and $M_\odot$ and $R_\odot$ are the solar mass and radius 
respectively, $u_{wsa}$ is the speed obtained by the WSA model and $p_0$, $\rho_0$, and $\gamma_0$ 
are the surface values of the pressure, density and $\gamma$ respectively.  Since the boundary 
conditions for the pressure and density are assumed to be known, we can obtain the boundary conditions for 
$\gamma$. In the next step, we introduce a radial function of $\gamma$ so its value changes from the surface 
value to a spherically uniform value of $5/3$ at $r=12.5R_\odot$. Finally, we solve the MHD equations self-consistently 
with non-uniform $\gamma$ until the wind solution converges.

We stress that this wind solution is determined by the magnetic field 
configuration. Therefore, any change in the field topology should change the wind 
structure in the same way that the magnetic field dictates the general structure of the coronal solution.


\section{SIMULATION}
\label{sec:Simulation}

In this work, we focus on the effect of star spot latitude on the angular momentum loss through the stellar wind.  
We have examined five test cases with different surface distributions of the magnetic field and have calculated the 
three-dimensional steady state solution for each case. The first case is a $5G$ dipolar field with no spots at all. 
The other cases are a $5G$ dipolar field superimposed with five spot pairs on each hemisphere, where the spot pairs 
are located at $\phi_s=36, 108, 180, 252, 324$ degrees in longitude. In each test case, we modify the spots co-latitude 
to be $\theta_s=10, 30, 60, 75$ degrees. For each case we use a spot magnetic field of $1kG$ and then repeat the 
simulation using a $200G$ field. The general description of the magnetic field for each latitudinal case can be written as: 
\begin{equation} 
B(\theta,\phi)=B_d+\sum_{\phi_s} A\cdot e^{-\left[ (\phi-\phi_s)^2+(\theta-\theta_s)^2 \right]/2.5^2\cos{\theta}}, 
\end{equation} 
where $B_d$ is the $5G$ dipolar field, $A$ is the field strength required for a magnetic flux of $1kG$ (or $200G$) 
in a spot area of $6.25\cos{\theta}$ degrees, and the summation is over the longitudes, $\phi_s$, for a particular 
latitude, $\theta_s$. $\theta$ and $\phi$ are both in units of degrees. The magnetic flux here
corresponds to the quantity that would be obtained from stellar magnetogram observations using Zeeman-Doppler imaging.

The polarity of each spot pair compared to the large-scale dipole is matched to the observed polarity 
of the emerging flux on the Sun \citep{babcock61}. This perfect alignment of the active region is probably 
not realistic, since differential rotation will cause one side of the spot to trail after the other one. 
Therefore, our three-dimensional simulation could be thought as a series of two-dimensional axial simulations.

\begin{figure*}[h!]
\centering
\includegraphics[width=6in]{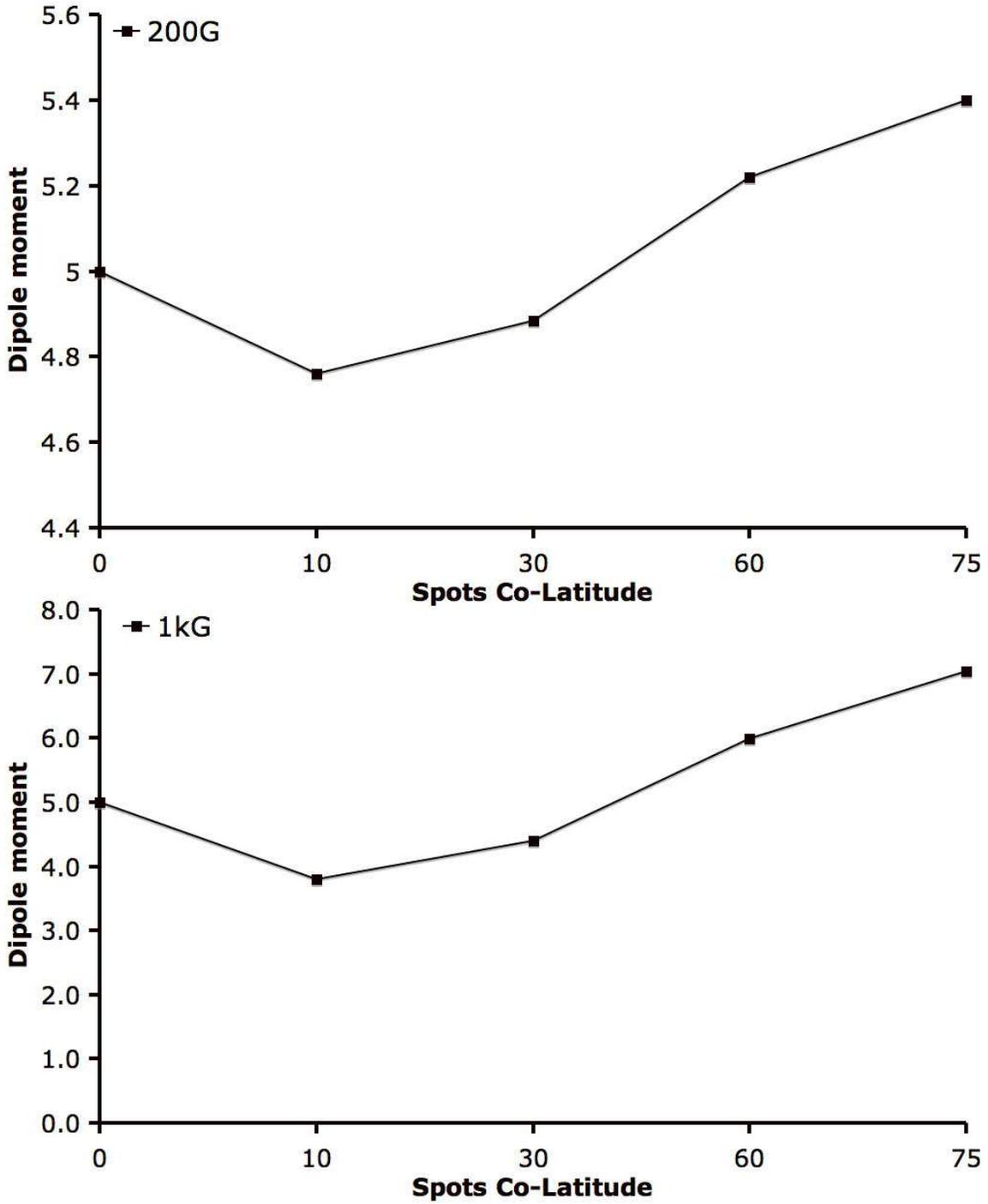}
\caption{The $g_1^0$ (dipole) component of the magnetic field harmonic expansion with spot strength 
of $200G$ (top) amd $1kG$ (bottom) as a function of co-latitude.}
\label{fig:f1}
\end{figure*}

In each case, the artificial magnetogram and potential field are calculated. Table \ref{table:1} and Figure \ref{fig:f1} 
show the $g_1^0$ (dipole) components for each case. It is clear that the dipole moment gets weaker for low latitude spots 
(less than 30 degrees) while it gets stronger with high latitude spots. The potential field distribution serves as the initial 
condition for the magnetic field in our simulation.  We then calculate the WSA speed distribution and the distribution of $\gamma$ 
for each case. Finally, we run the MHD code until a steady-state is obtained.

In all test cases, initial and boundary condition are determined in the manner described by
\cite{cohen07a}. We use a Cartesian grid with 10 levels of mesh refinement so that the grid cell size 
near the inner boundary is of the order of $1/200\;R_\star$, where $R_\star$ is the stellar radius. This high 
resolution is required in order to resolve the active region with at least several grid cells. We also 
dynamically refine the grid around current sheets that appear in the simulation. All MHD computations 
were performed using the Pleiades cluster at NASA's NAS center.

\subsection{Wind dependency on the Magnetic Field Structure}
\label{WindDependency}

\begin{figure*}[t!]
\centering
\includegraphics[width=3in]{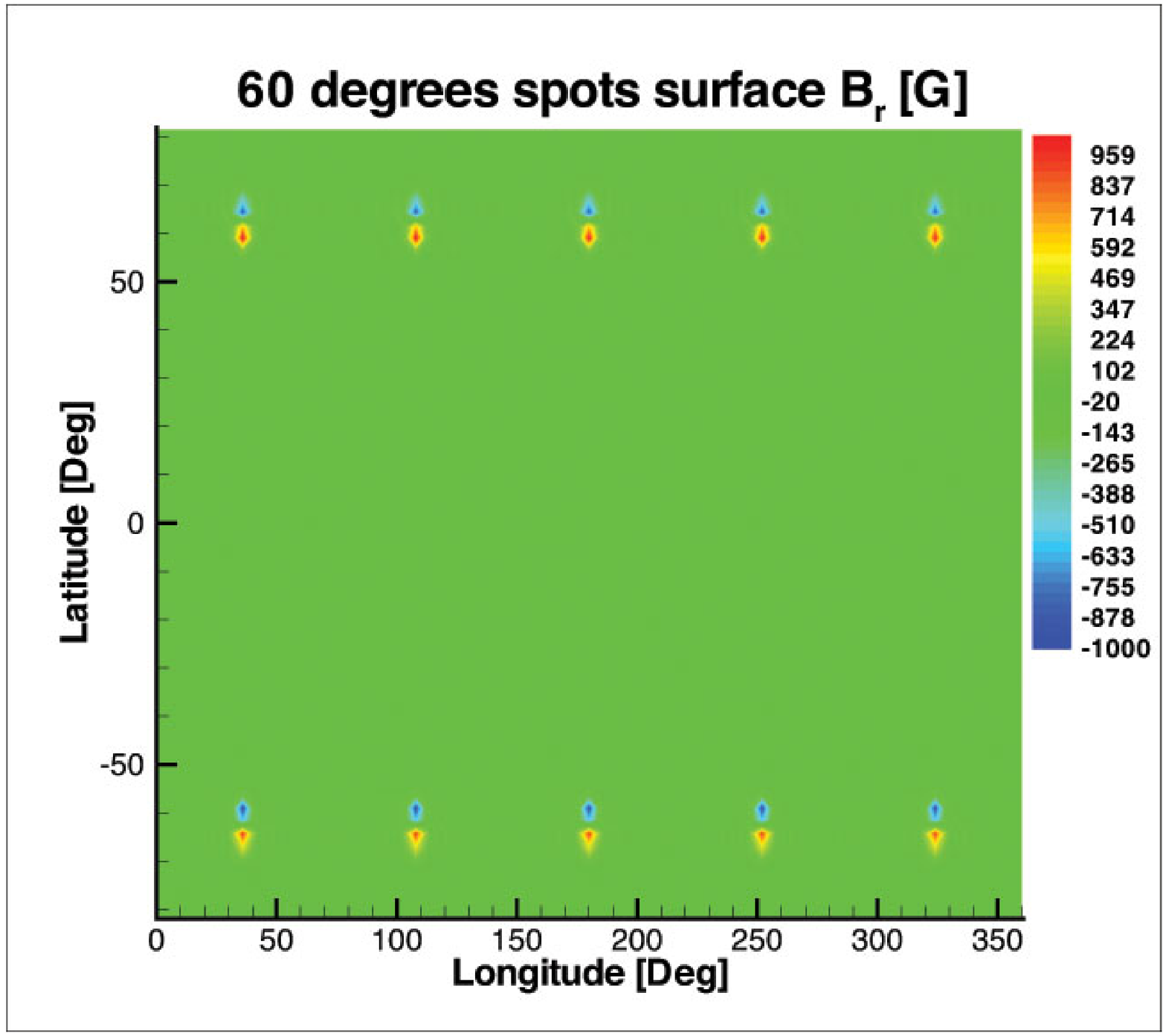}
\includegraphics[width=3in]{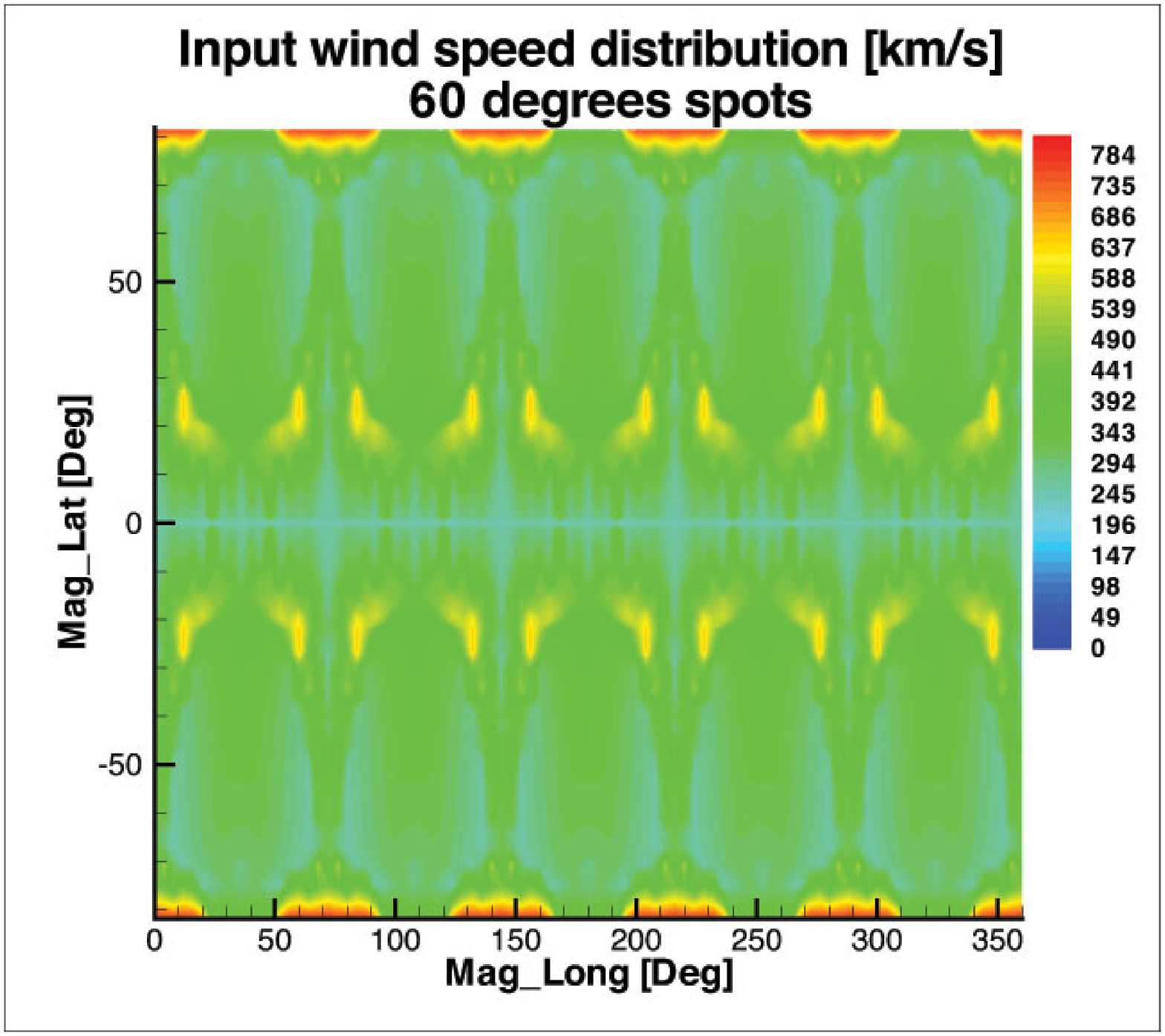} \\
\includegraphics[width=3in]{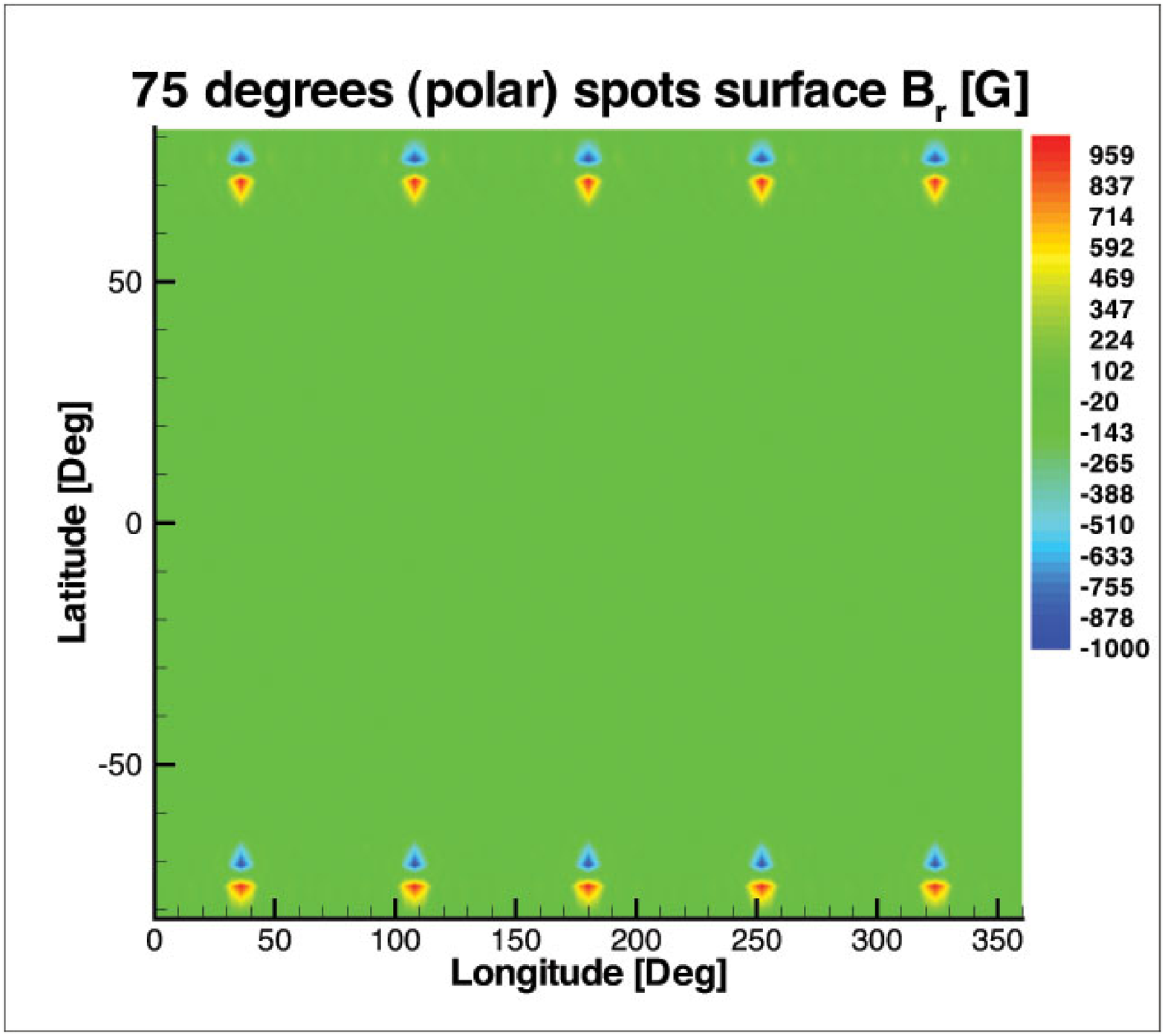}
\includegraphics[width=3in]{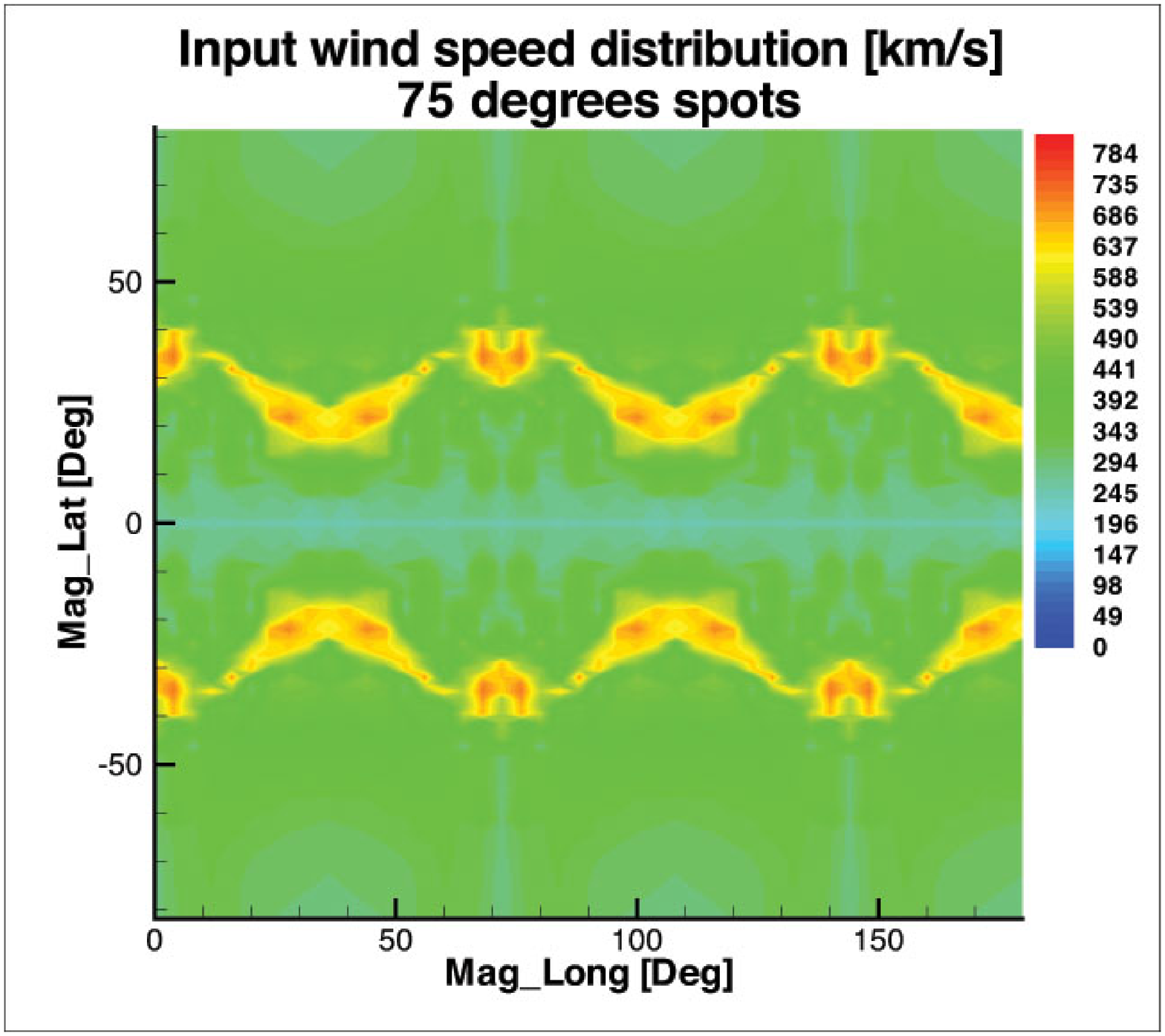} 
\caption{Surface radial field strength with color contours representing 
$B_r\;[G]$ (left) and resulted input speed distribution calculated by the WSA model with 
color contours representing $u_{wsa}\;[km\;s^{-1}]$ (right). The plots are for (top to bottom) 
dipolar field with no spots, and spots at 10, 30, 60, 75 degrees from the equator. 
Figures 1a-1j are available in the online version of the Journal.}
\label{fig:f2}
\end{figure*}

Figure~\ref{fig:f2} shows the surface radial magnetic field strength for each test case (left) 
and the resulting wind speed distribution calculated from the WSA model (right). 
It is obvious here that there is a strong dependency of the wind structure 
on the magnetic field topology. The main effect in the case of introducing strong spots on top of a 
dipolar field is that the fast wind gets eliminated as the spots migrate towards the pole. In 
the polar spot case, there is no fast wind at all except for two strips of moderate wind ($500\;km\;s^{-1}$) at 
mid-latitudes, which is consistent with the study of \cite{mcIvor04}.

\begin{figure*}[h]
\centering
\includegraphics[width=3in]{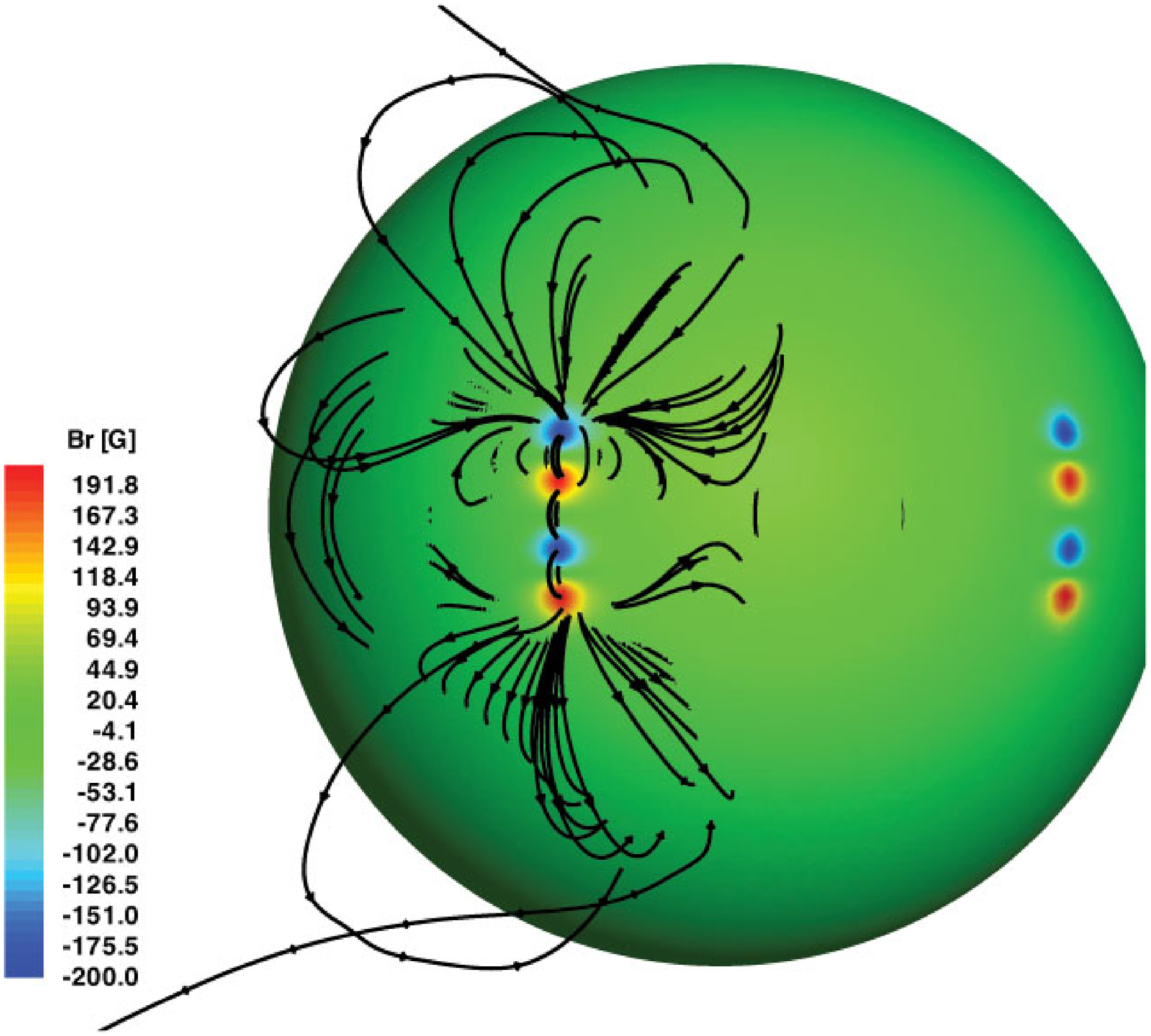}
\includegraphics[width=3in]{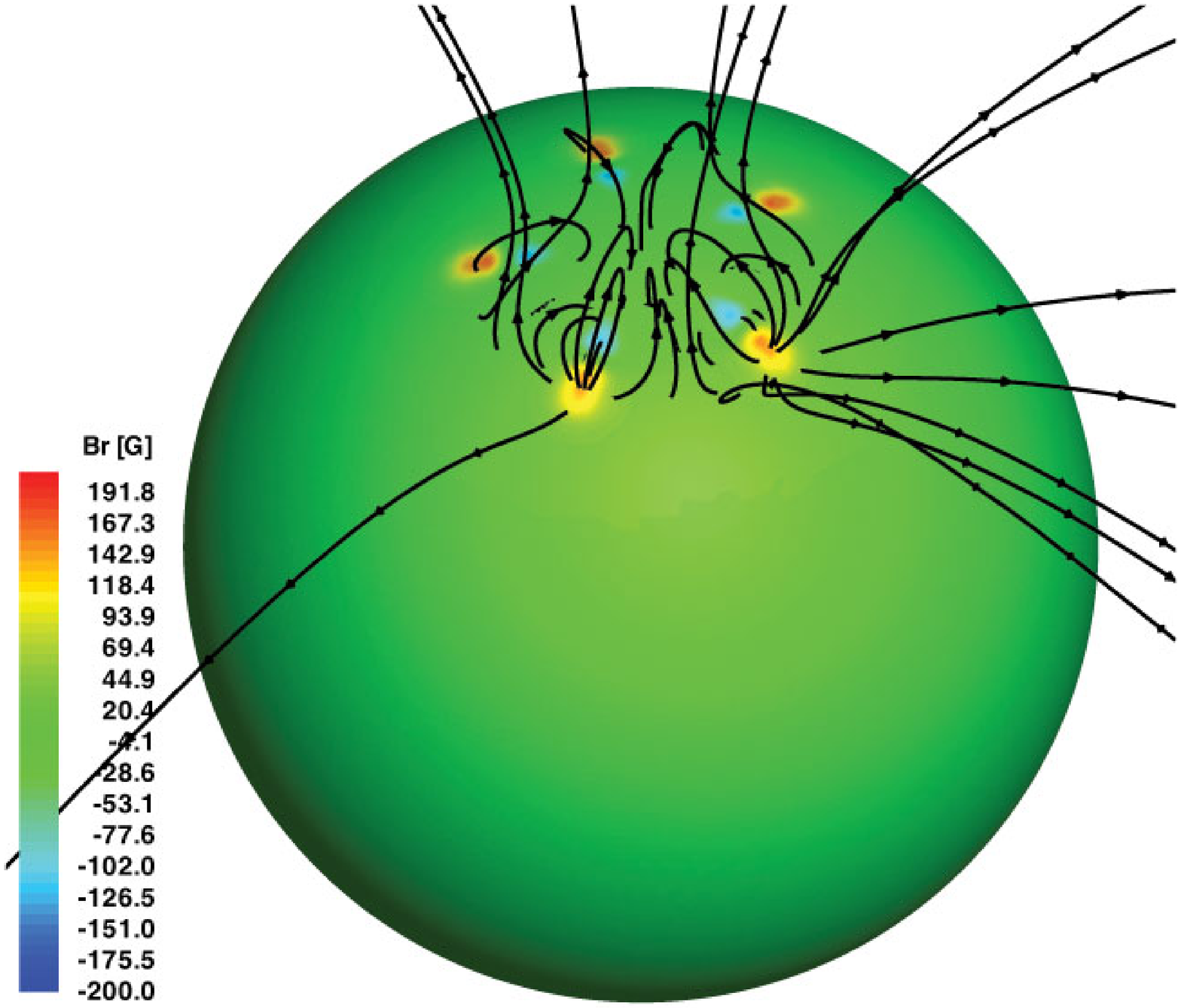}
\caption{Star spots located at $10^\circ$ from the equator (left) interact mostly with the 
closed field streamers, while polar spots located at $80^\circ$ (right) interact 
mostly with open flux. The sphere is of $r=1.04R_\star$ so $B_r$ of the spots is weaker than the surface value.
Figures 2a-2b are available in the online version of the Journal.}
\label{fig:f3}
\end{figure*}

The reason for this behavior can be found by observing the detailed structure of the magnetic field in 
the vicinity of a star spot. Figure~\ref{fig:f3} shows 
the magnetic field topology near the star spot for the equatorial case (left) and the polar 
case (right). The surface is at $r=1.04R_\star$, color contours are of radial field strength and streamlines 
represent the three-dimensional magnetic field lines. The fast wind comes from open field regions, where the 
expansion of the flux tubes is small \citep{wangy90}. In the equatorial case, the star spots interact mostly with the closed 
streamers and only slightly with open flux located at the coronal hole boundary. Therefore, there will be some 
closing of open field lines and a slight increase in the expansion of the tube, resulting in some removal 
of the fast wind. In the polar case, the star spots interact completely with field lines originally open, so 
that there are no longer regions with non-expanding flux tubes and the fast wind is completely 
eliminated.  The regions with the minimum expansion are located at mid-latitude where we find 
moderate wind strips.

\subsection{Mass and Angular Momentum Loss Rates}
\label{MdotJdot}

Once a steady state is obtained for each case, we calculate the Alfv\'enic Mach number, $M_A=u_{sw}/u_A$, 
with $u_A=B/\sqrt{4\pi \rho}$ being the Alfv\'en speed, and compute the Alfv\'en surface as that where $M_A=1$. 
The mass loss rate through the wind, $\dot{M}$, can be calculated by integrating the mass flux through the Alfv\'en surface:
\begin{equation}
\dot{M}=\int \rho \mathbf{u}\cdot\mathbf{da}_A,
\end{equation}      
where $\mathbf{da}_A$ represents a surface element on the Alfv\'en surface. The mass flux does not depend of course 
on the particular surface of integration, since it is conserved. 

In a similar manner, the AML rate, $\dot{J}$, can be obtained using Eq.~\ref{WD}:
\begin{equation}
\dot{J}=\frac{3}{2}\int \Omega \sin{\theta} r^2_{A} \rho \mathbf{u}\cdot\mathbf{da}_A,
\end{equation}
where $r_{A}$ and $\theta$ are the local radius and latitude at each element 
of the Alfv\'en surface.

The accuracy of the results depends mostly on the grid resolution. Since we use a non-uniform grid, 
we estimate the accuracy of our calculation for the largest grid cells on the Alfv\'en surface, assuming 
the result would be more accurate were smaller grid cells used. 
We found that for the largest grid cells 
$\Delta\dot{M}/\dot{M} \le 0.001$ and $\Delta\dot{J}/\dot{J} \le 0.1$.

\subsection{Results for the Case of 1kG Spots}
\label{Results1kG}

\begin{figure*}[t!]
\centering
\includegraphics[width=3in]{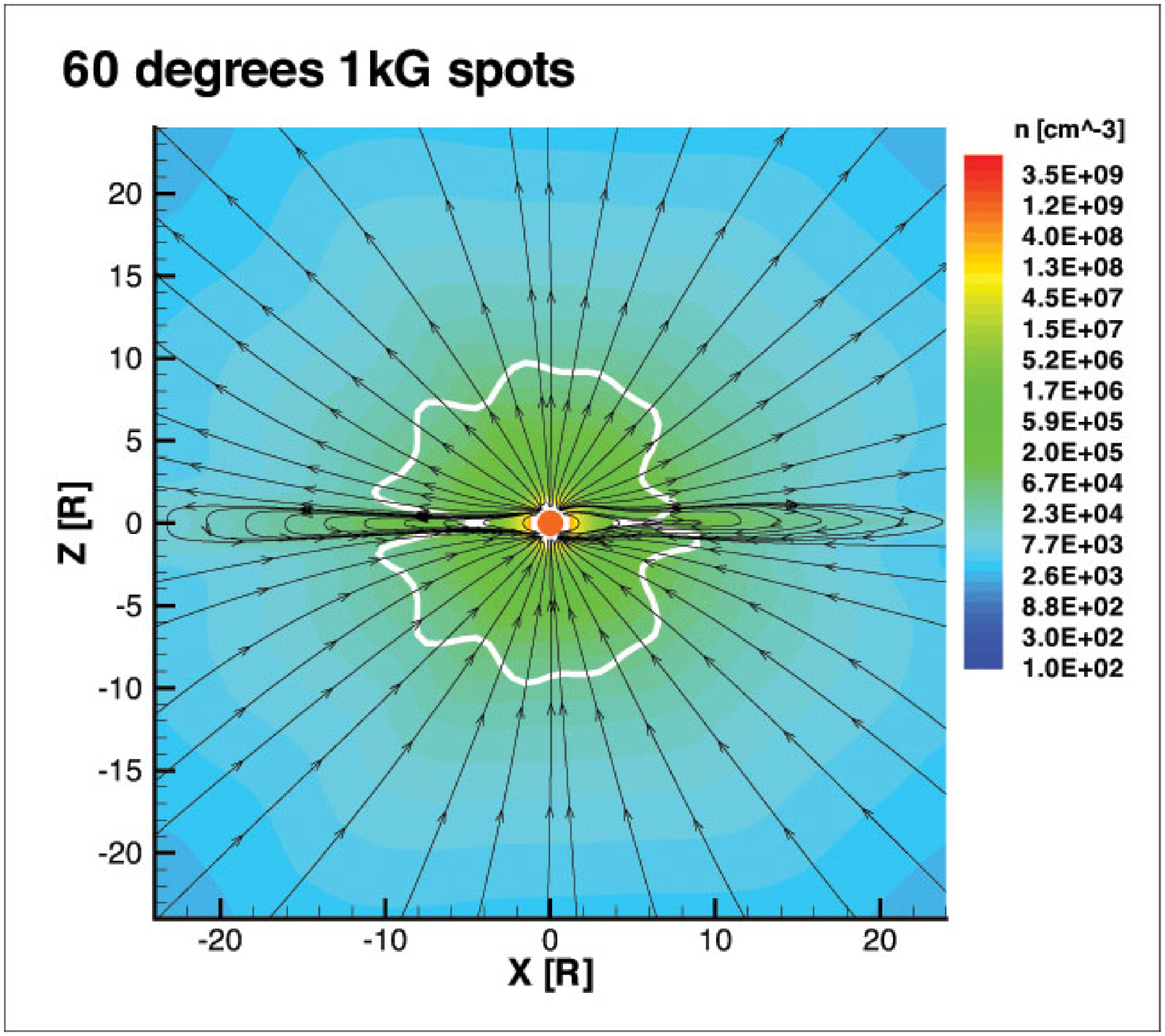}
\includegraphics[width=3in]{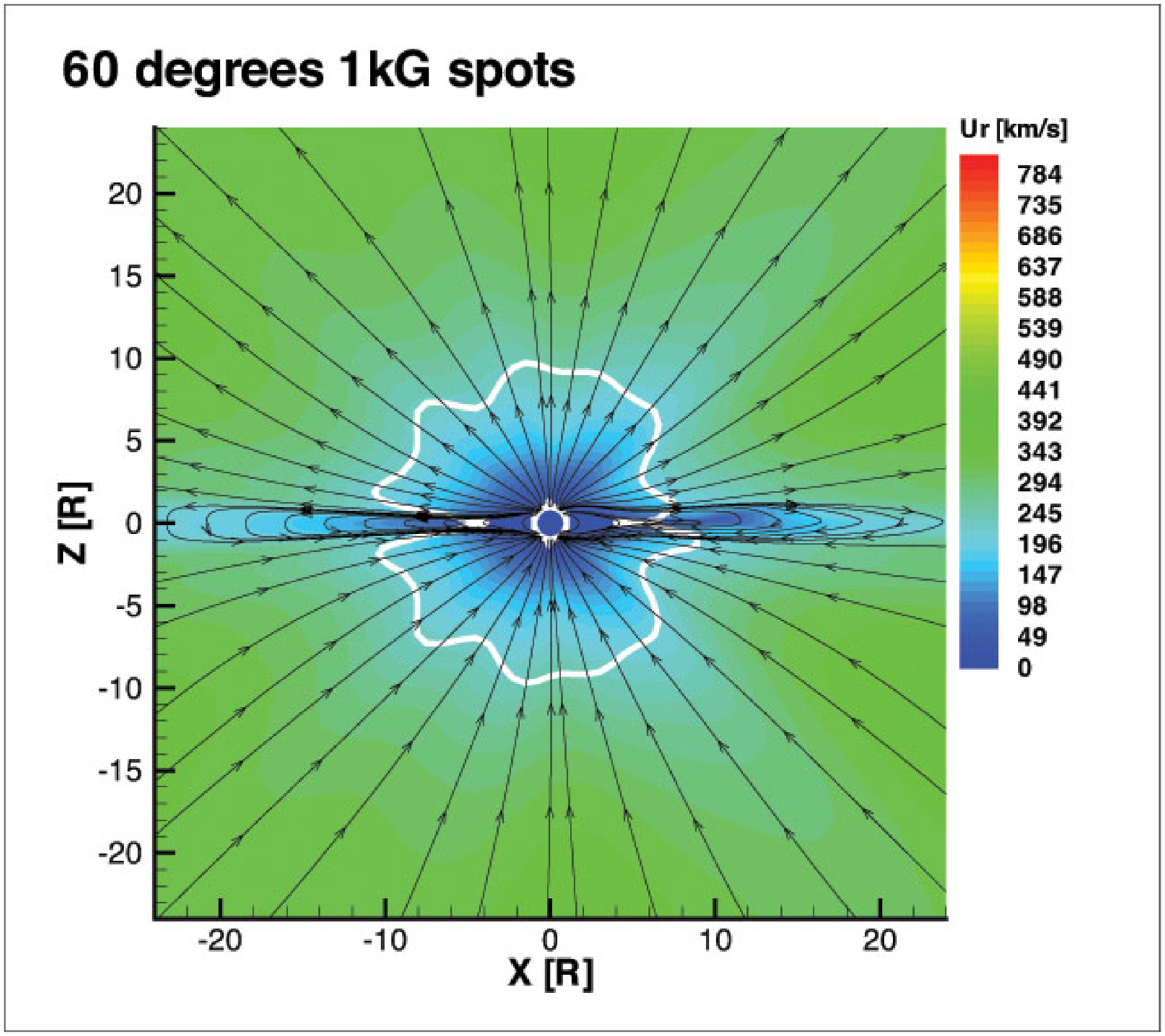} \\
\includegraphics[width=3in]{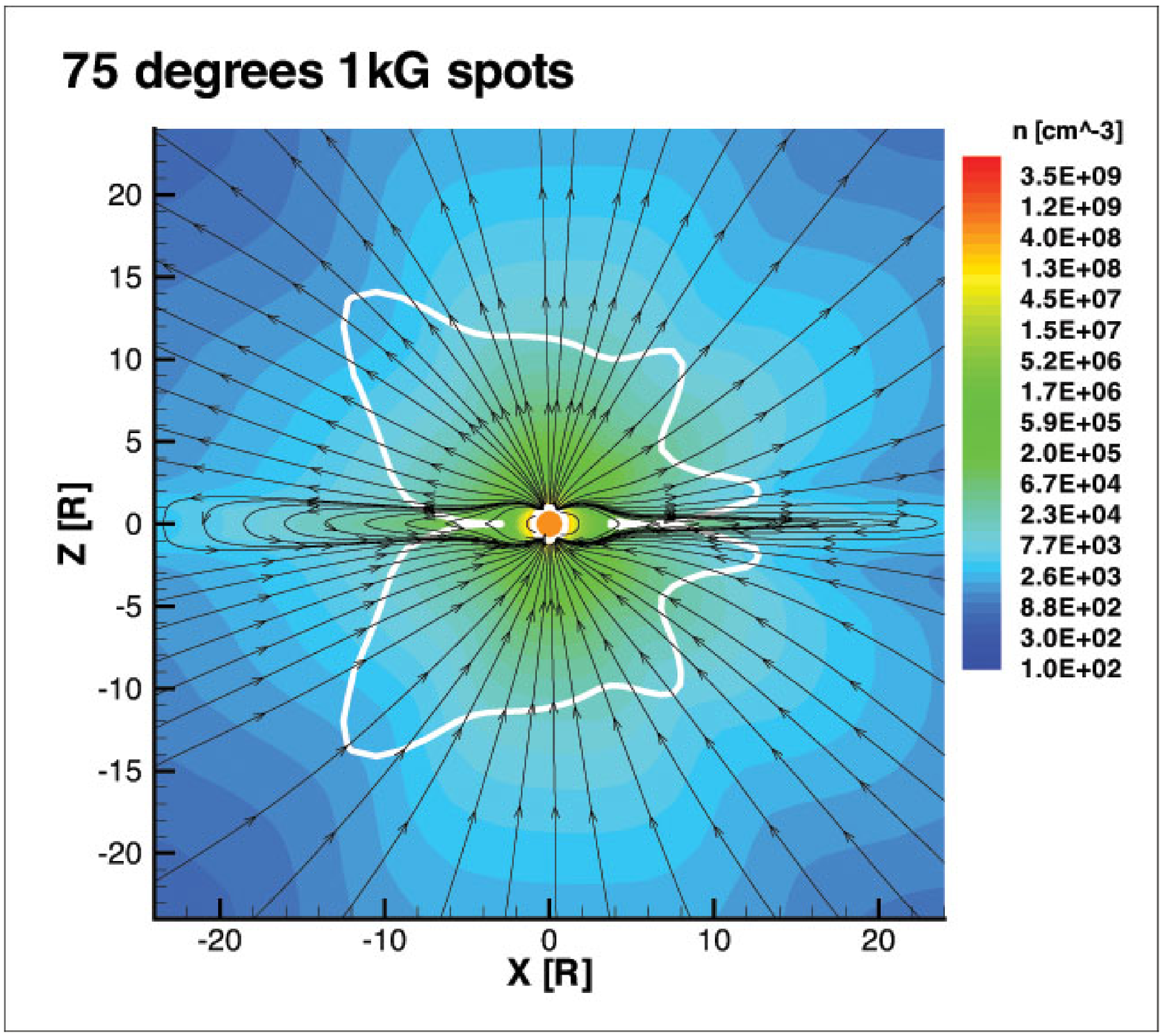}
\includegraphics[width=3in]{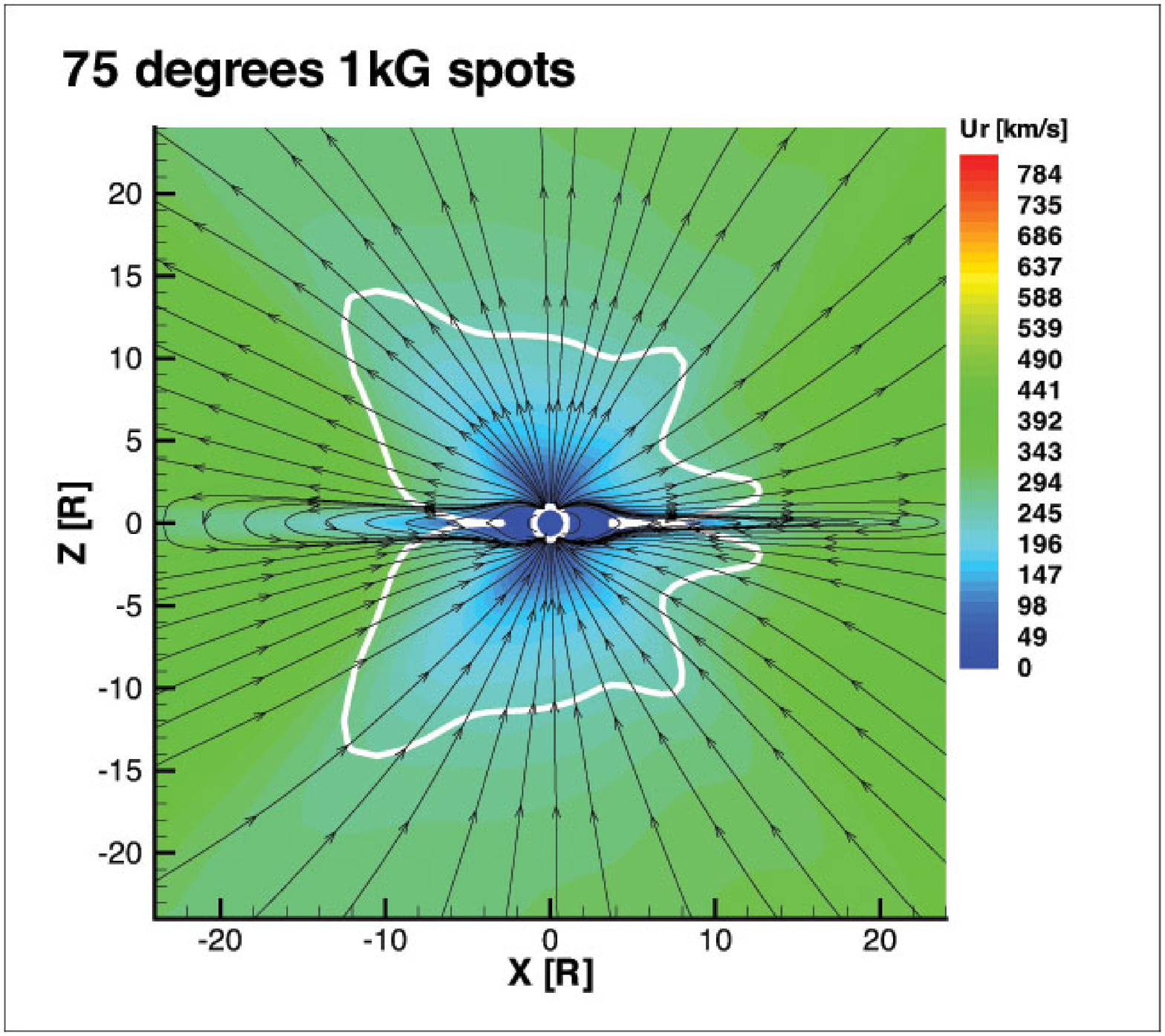}
\caption{A y=0 cuts in the three-dimensional steady state solutions contain $1kG$ spots. Top to bottom 
show the no spots case and the cases in which spots are located at co-latitude of 10, 30, 60, and 75 
degrees. Color contours are of number density (left) and radial wind speed (right). Streamlines represent 
the magnetic field lines, the white solid line represents the Alfv\'en surface.Figures 3a-3j are available 
in the online version of the Journal.}
\label{fig:f4}
\end{figure*}

Figure~\ref{fig:f4} shows the $y=0$ plane from the three-dimensional steady state solution for 
each test case (top to bottom). The color contours are of number density (left) and 
radial wind speed (right). Black streamlines in each panel represent the magnetic field lines, 
while the solid white line represents the Alfv\'en surface. The left-right asymmetry is due to 
the fact that the cut goes through the spots on the left side, while it goes between two spots on 
the right side.

The existence of the strong 
magnetic field from the spots locates the Alfv\'en surface in this direction further 
out at larger radial distance than in the 
pure dipolar field case.  Another consequence of the increase in magnetic field strength due to the strong 
spots is that the current sheet is wider than in the dipolar case. 

The results of the MHD solution are consistent with the input speed distribution obtained by the WSA model, 
and the figures shows how the fast wind is being eliminated as the spots migrate towards the pole. 
As the spots move towards the pole, 
the Alfv\'en surface shrinks and is smallest for the case of 60 degrees co-latitude. The Alfv\'en surface gets 
bigger for the polar case, since the field strength at high latitudes on the surface is higher  
due to the polar spots. In addition, the overall slowdown in the wind for 
more poleward spots causes the density to increase, so 
we find a significant increase in the total mass within the co-rotating volume between the surface 
of the star and the Alfv\'en surface.  

\subsection{Results for the Case of 200G Spots}
\label{Results200G}

\begin{figure*}[t!]
\centering
\includegraphics[width=3in]{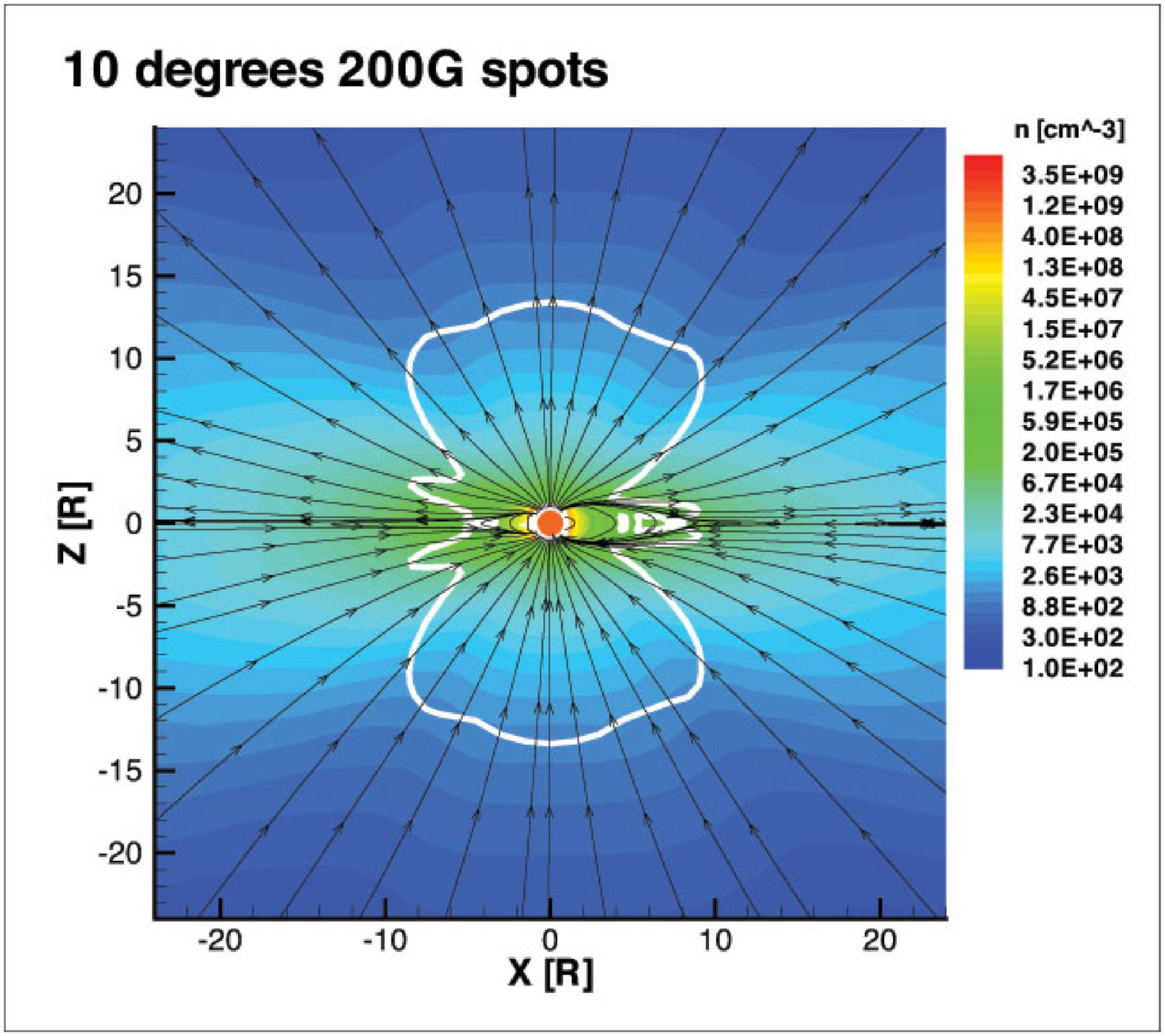}
\includegraphics[width=3in]{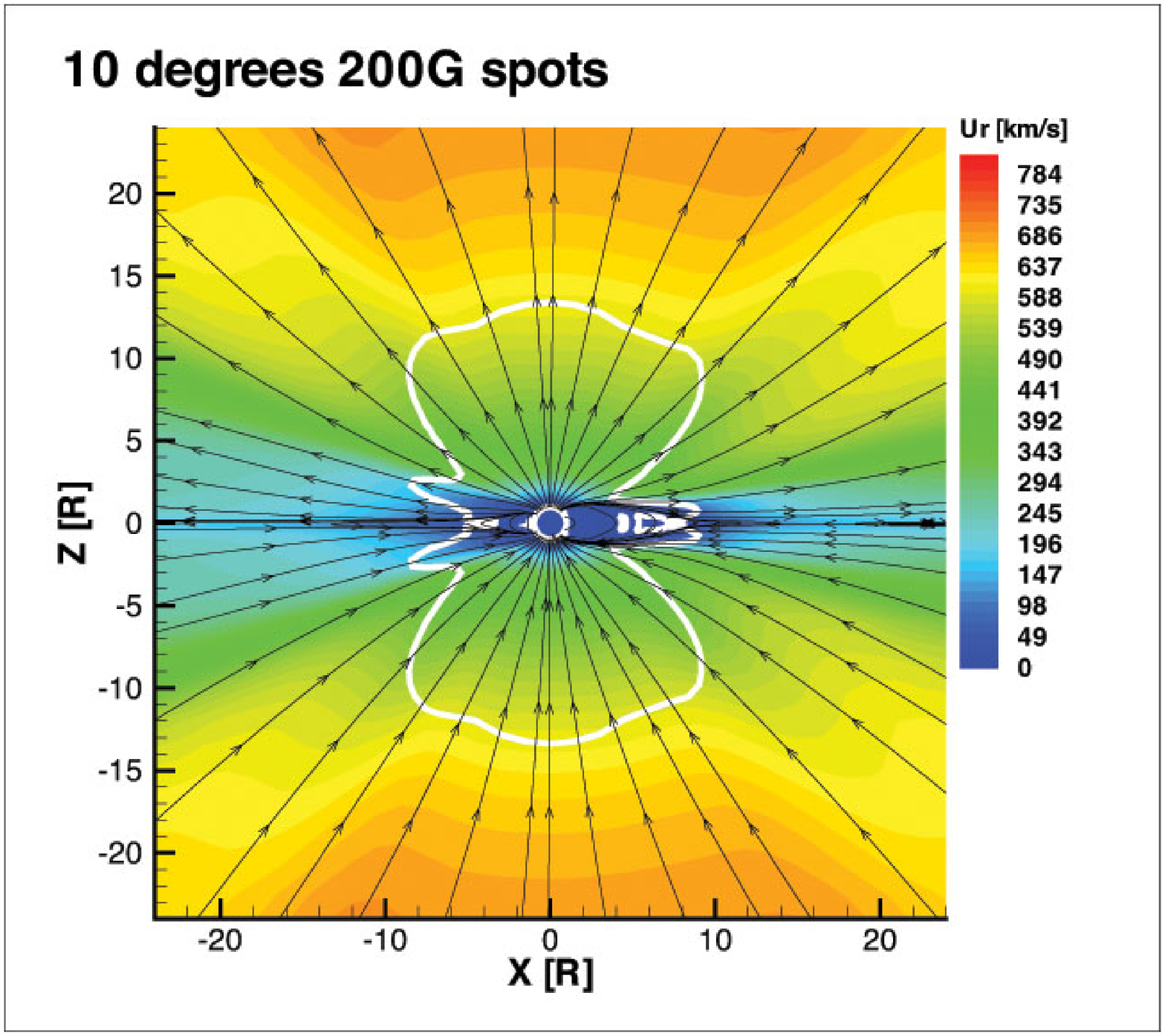} \\
\includegraphics[width=3in]{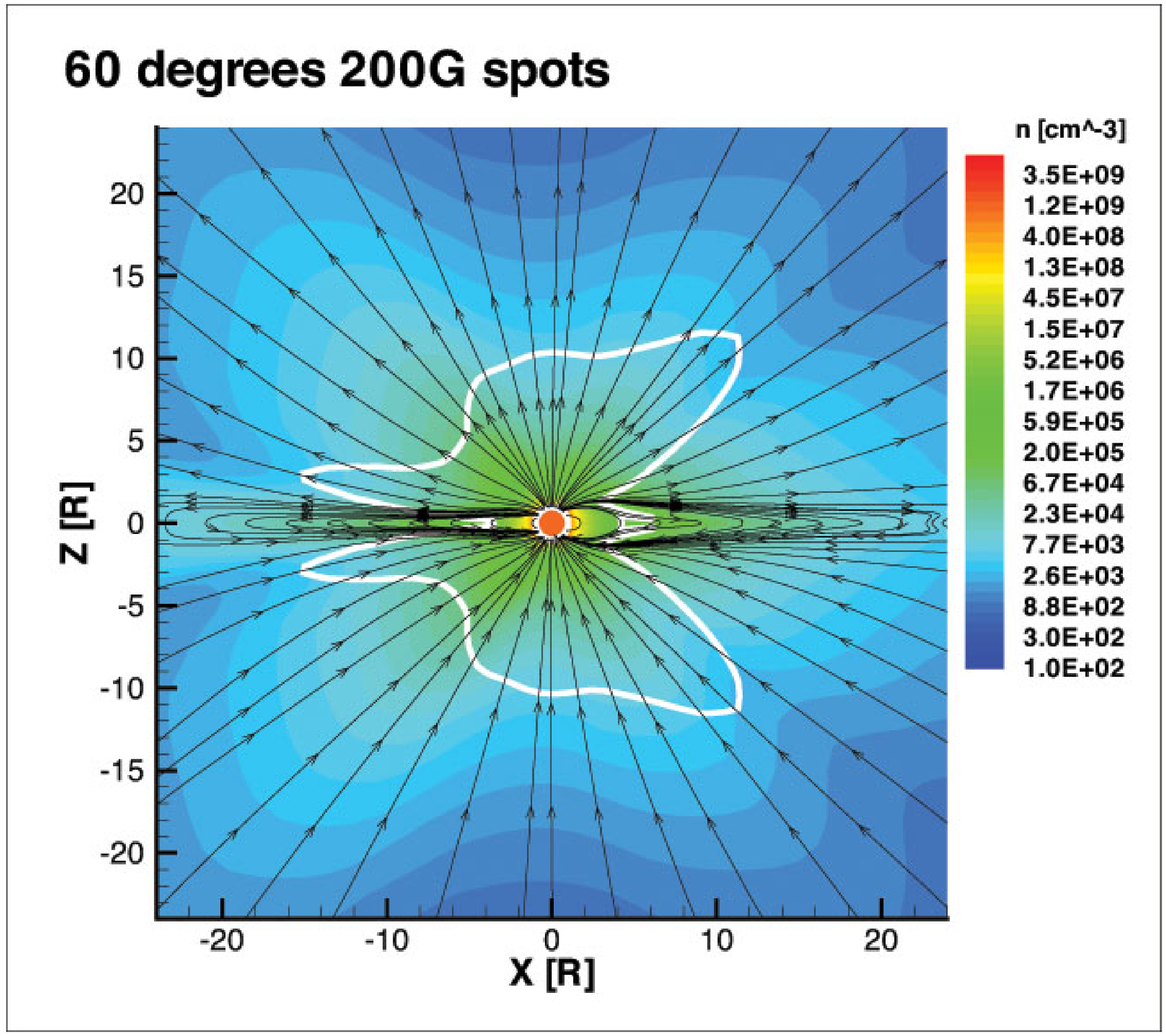}
\includegraphics[width=3in]{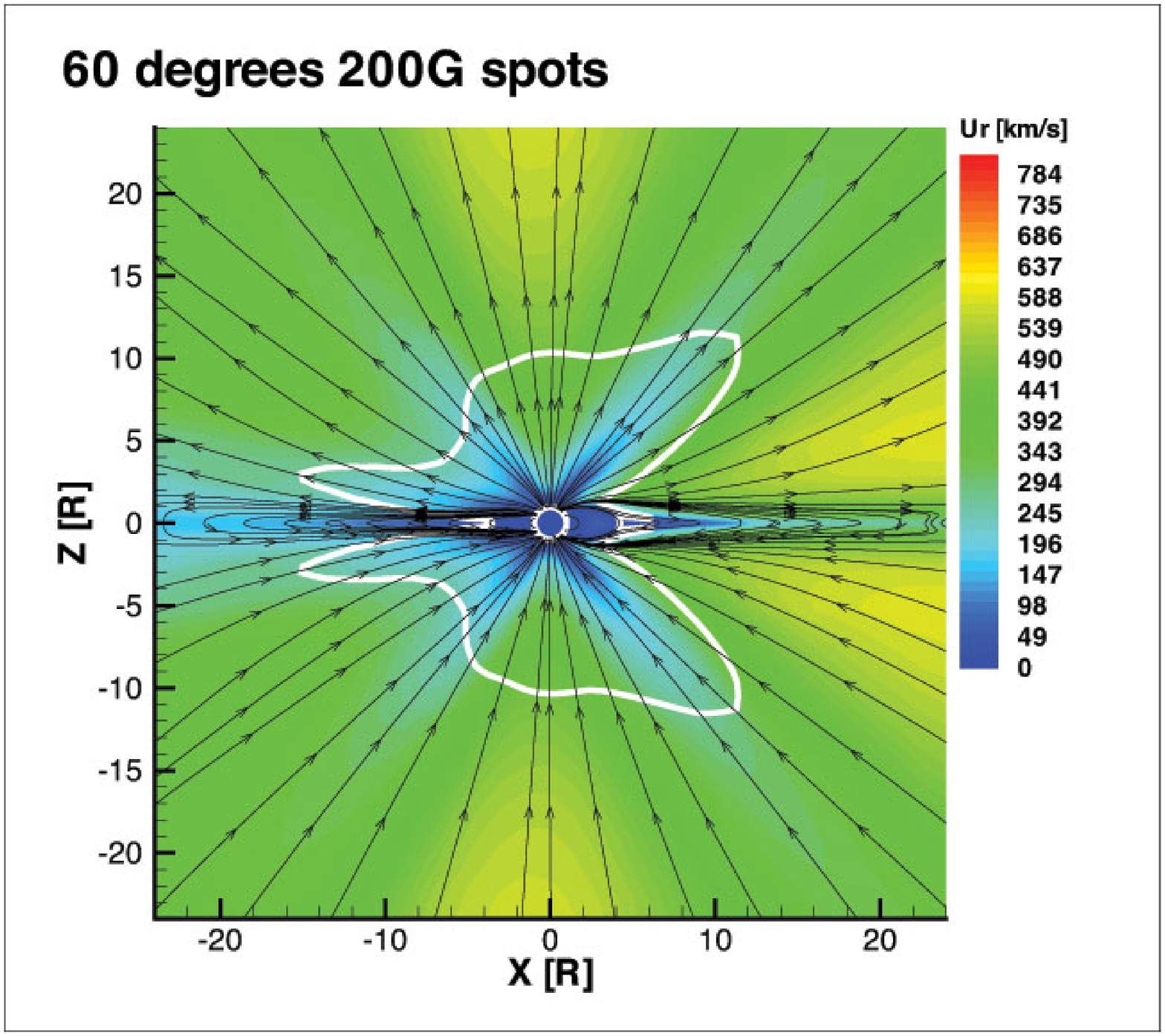} \\
\includegraphics[width=3in]{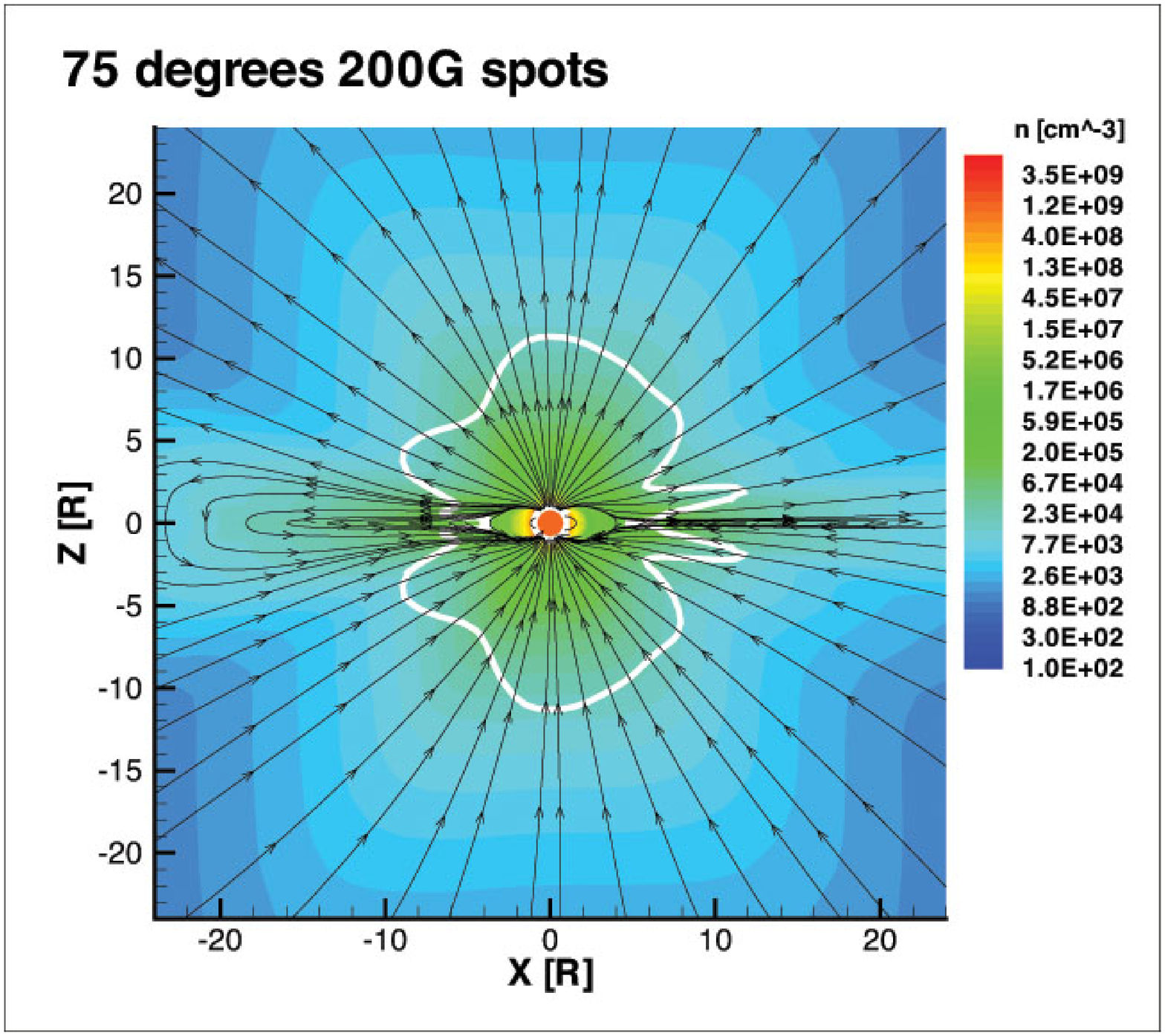}
\includegraphics[width=3in]{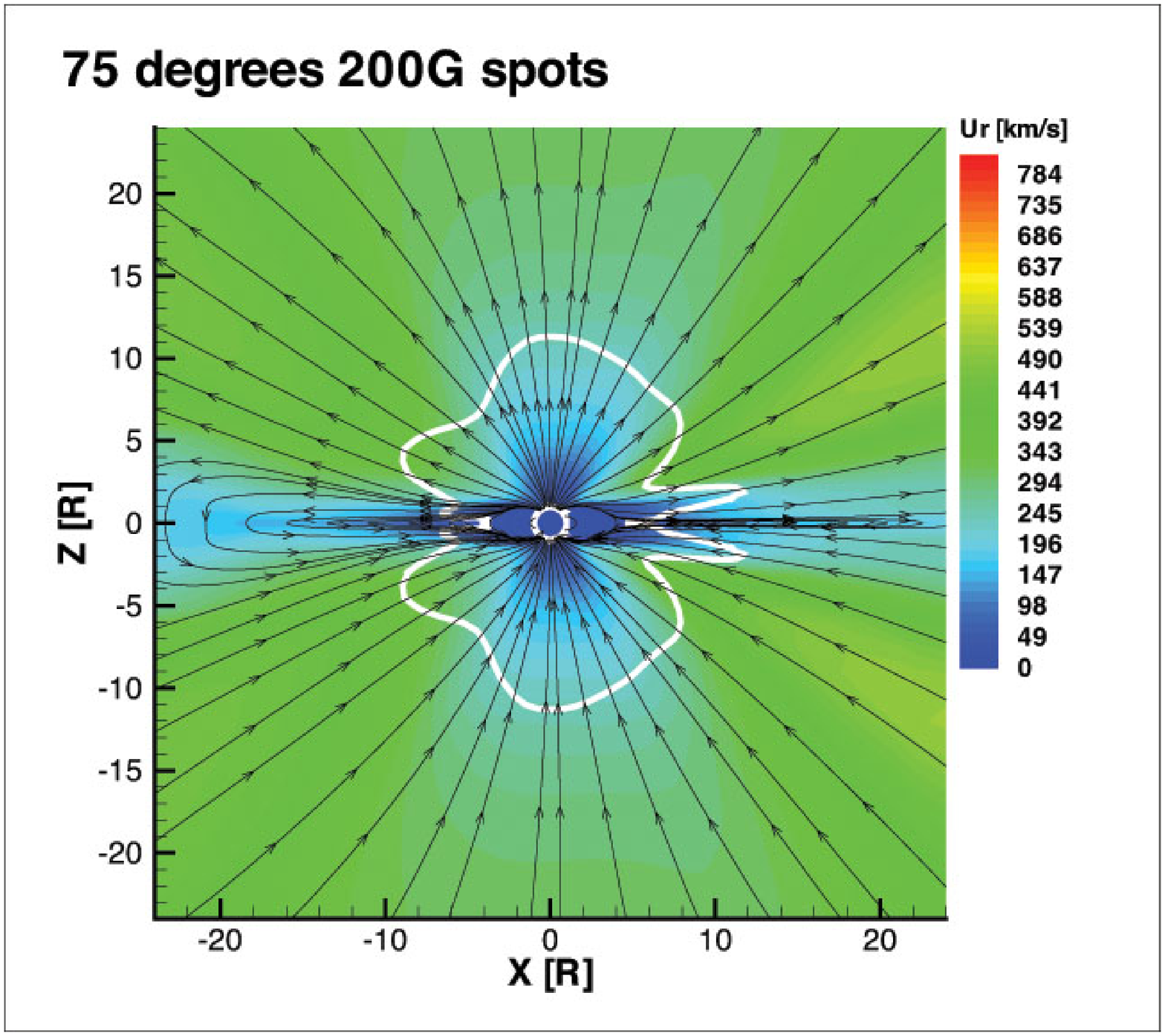}
\caption{A y=0 cuts in the three-dimensional steady state solutions contain 200G spots. Top to bottom show the cases 
in which spots are located at co-latitude of 10, 60, and 75 degrees. Display is the same as Figure~\ref{fig:f3}; 
figures 4a-4f are available in the online version of the Journal.}
\label{fig:f5}
\end{figure*}

Figure~\ref{fig:f5} shows the simulation results with spot magnetic fields of $200G$. It can be seen for 
this weaker field that the effect of fast wind elimination is reduced  due to the correspondingly 
weaker interaction of the spot and ambient fields. Unlike the $1kG$ case for 60 and 75 degrees, 
in which wind over $300-400\;km\;s^{-1}$ has been completely eliminated, in the 200G case there is 
still some faster wind at higher latitudes.


\section{DISCUSSION}
\label{sec:Discussion}

\subsection{The Effect of Wind Distribution on AML}

\begin{figure*}[t!]
\centering
\includegraphics[width=6in]{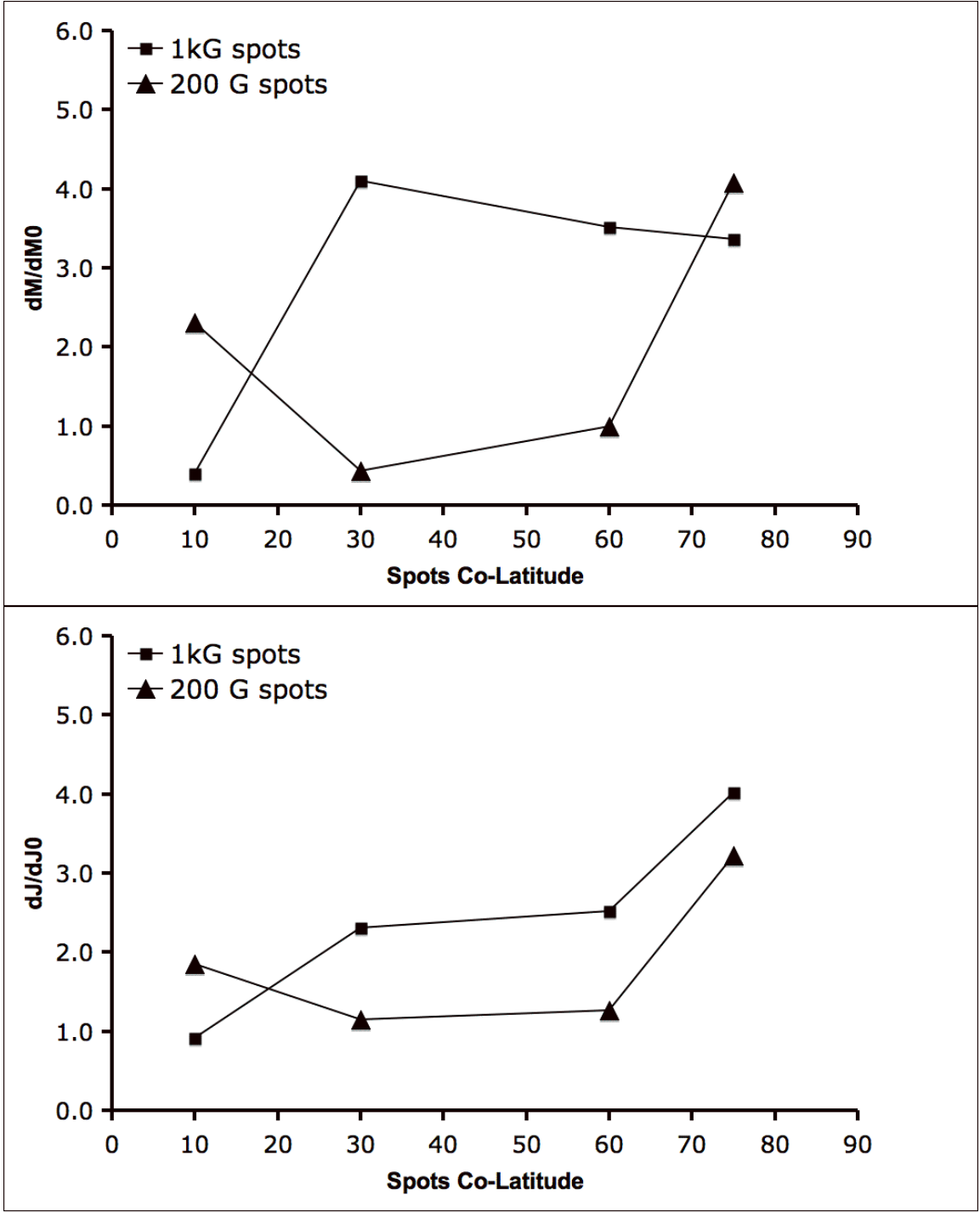}
\caption{Line plot of mass (top) and angular momentum (bottom) loss rates denoted by $dM$ and $dJ$ as a function of 
spots co-latitude for $1kG$ spots (squares) and 200G spots (triangles). Rates are normalized to the loss rates 
of the dipolar case $dM_0$ and $dJ_0$.}
\label{fig:f6}
\end{figure*}

Figure~\ref{fig:f6} shows line plots of mass loss rate (top) and angular momentum loss rate (bottom), 
denoted by $dM$ and $dJ$, as a function of spot co-latitude for the $1kG$ case (squares) and the 
200G case (triangles). The rates are normalized to the mass loss rate 
$dM_0=1.5\times 10^{-14}\;M_\odot \; yr^{-1}$, and the AML rate $dJ_0=10^{30}\;g\;cm^2\;s^{-2}$, of the dipolar case. 
$dM_0$ is also consistent with the solar mass loss rate calculated using typical solar wind parameters 
at 1~AU \citep[e.g.][]{gombosibook}.

The loss rates depend on the interplay between the wind speed, density, and 
the size of the Alfv\'en surface. In principle, one should expect the 
AML rate to be controlled mainly by the mass flux. However, this seems to be true 
only when the fast wind dominates, as in the dipolar case or when the spots are near the equator.  

In the dipole field case with no spots, the rates depend on the high mass flux due to the 
fast wind at high latitudes, and on the relatively high density and stronger magnetic field 
close to the equator. When introducing equatorial spots, they interact with the closed field 
streamers so the density structure near the current sheet does not change much. In this case, 
the equatorial effects on the loss rates are not significantly different and the 
AML rate is controlled by the change in the mass flux.

\begin{figure*}[t!]
\centering
\includegraphics[width=6in]{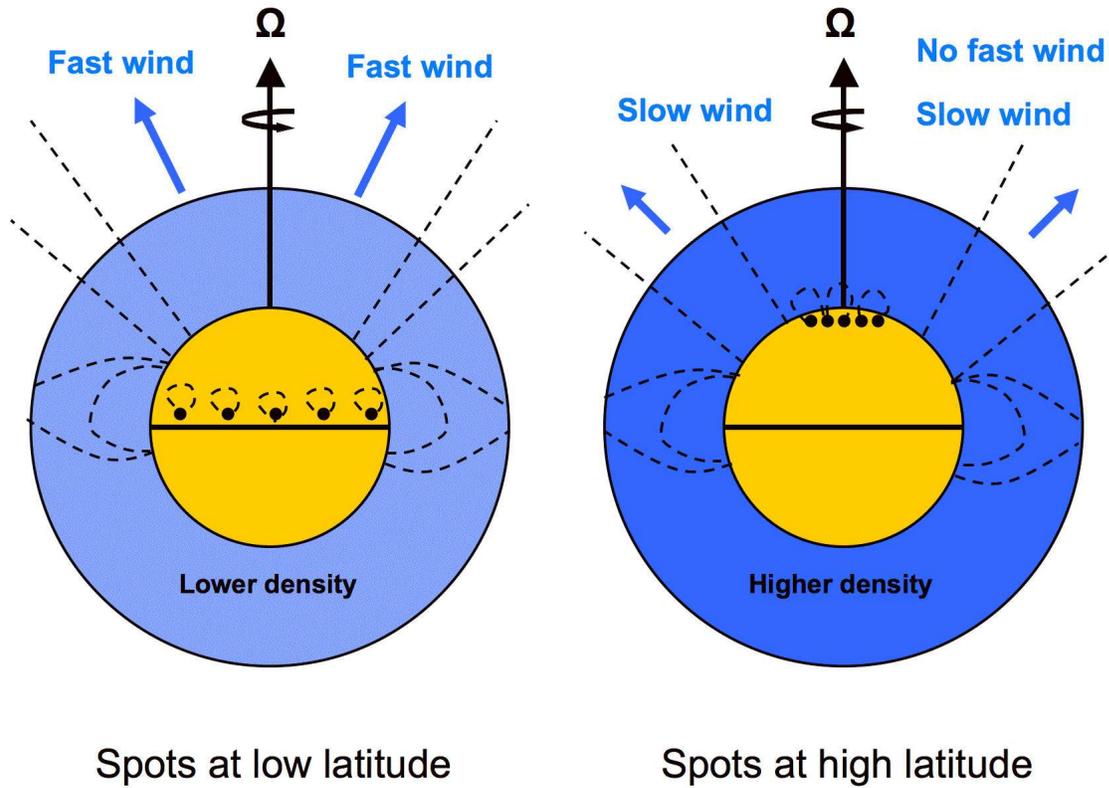}
\caption{The effect of spot location on the AML. The inner circle represents the star and the 
outer circle represents the Alfv\'en surface with tentative field lines (dashed lines). The left panel shows the case of low-latitude spots, in which the AML 
is controlled by the high mass loss through the fast wind at high latitudes. In the right panel, which shows the 
situation for high-latitude strong spots, the fast wind is eliminated and so does the high mass loss at high latitudes. 
As a result, the total density in the volume is high and applies more torque on the star due to the imperfect co-rotation 
and the fact that it is frozen-in to the magnetic field.}
\label{fig:f7}
\end{figure*}

When the spots are located at higher latitudes the amount of fast wind is 
decreased and the AML is actually affected most by the significant increase in 
density in the volume between the stellar surface and the Alfv\'en surface.
It is possible that this high density, slow wind is related to the accumulation
of H$\alpha$-emitting gas in ``slingshot'' prominences now commonly observed
on rapidly rotating stars \citep{colliercameron89,colliercameron92,jeffries93,
byrne96,eibe98,donati00,jardine05}. The star tries to co-rotate the plasma that 
exists between its surface and the Alfv\'en surface, but the stronger field and 
higher plasma mass density apply more torque on the star. The net result is higher 
loss of angular momentum, which in this case might be described as AML to the 
stellar outer atmosphere rather to the stellar wind. The mass flux is higher for 
the case of polar spots with 200G than for just the dipole (as seen at the top of 
Figure~\ref{fig:f6}), and the net AML is still greater for the case of $1kG$ spots 
(as seen at the bottom of Figure~\ref{fig:f6}). Figure~\ref{fig:f7} shows a cartoon of 
the effect of spot location on AML.

The overall trend in Figure~\ref{fig:f5} inverts below 10 degrees due to the size of the Alfv\'en surface. 
In the case of $1kG$ spots, more fast wind is eliminated and the Alfv\'en surface is expanded, so the total mass 
flux and AML are lower. In the case of $200G$ spots, the Alfv\'en surface is smaller than the case with no spots. 
While in this case the wind distribution is similar to the no spots case, the Alfv\'en surface is now located in 
regions of higher density so the total mass flux through the surface is higher.

\subsection{Implications for the Long Term Stellar Evolution}

The appearance of polar spots in fast rotating stars has been explained by the enhancement of 
Coriolis force acting on the emerging flux tubes \citep{schuesslersolanki92} or alternatively 
by the behavior of the meridional circulation in fast rotating stars \citep{schrijvertitle01}. 
In both cases the rotation rate affects the magnetic field distribution. \cite{aibeo07} have argued 
that re-distribution of coronal flux in stars does not significantly change the wind efficiency for 
the magnetic breaking. However, their model took into account only existing magnetic flux with no addition of 
new emerging flux (i.e. active regions or spots). Our results clearly show that the AML through the wind 
is affected by the location, as well as the strength, of the spots. Therefore, we argue that as a result of 
fast-rotating young stars having their new magnetic flux appear in high latitudes, their winds are organized 
in such a way that the AML and the spin down of the star are large. As the rotation rate decreases, 
the new magnetic flux appears at lower latitudes and consequently the wind is organized in a way that the AML 
decreases as well.

The aim of our study is to begin to understand how angular momentum loss changes with surface magnetic topology, 
with the ultimate aim of understanding the rotational evolution of stars. On more 
magnetically active stars, we expect the strength of the spots, as well as the strength of the background dipole, 
to be higher than the Sun-like case. The stronger surface field will induce a stronger stellar wind, so the 'ambient' 
mass-loss and AML rates from these stars would be higher, regardless of the magnetic spot distribution. Here, we have 
examined how redistributing the surface field affects the stellar wind distribution and its AML. The dominant effect 
we found was the re-organization of the open/closed field lines for the different spot latitudes, and elimination of 
the fast wind for high latitude spots. As the strength of the dipole is increased, the influence of the spots on the 
magnetic field distribution will be diminished. For a very strong background dipole an order of magnitude or more 
higher than considered here, the effect of the spots is likely to be quite small. It is not yet known how the dipolar 
field scales with stellar activity in general, and future theoretical investigation of the effects of the dipolar field 
strength on AML would be worthwhile.

\citet{kawaler88} developed a general expression for the rotational evolution of
low-mass stars based on the WD formulism (Eqn.~\ref{WD}) and an assumed linear scaling 
of radial and dipolar surface magnetic field strength with rotation rate, $B_0\propto \Omega$. 
This simple type of approach has proven  reasonably successful at matching stellar rotation velocities 
vs.~age for solar-like stars of Hyades age (600Myr) and older 
\citep[e.g.][]{kawaler88,krishnamurthi97,sills00}.  However, the relation fails for earlier ages 
because the predicted AML is too high to explain rapid rotators with periods of
one to several days in younger open cluster stars \citep[e.g.][]{stauffer87,barnes96,krishnamurthi97}.
To avoid this problem \citet{krishnamurthi97} introduced a "saturation" in the dependence of the surface 
magnetic field on rotation rate: beyond some critical period the surface field was assumed constant.

Rotation periods of one to several days also represents the regime in which high latitude
surface spots are observed in single and tidally-locked binaries \citep[e.g.][]{strassmeier01}. 
It is in this regime that the complications of spot latitude on AML discussed here will be important. 
The results indicate that a simple saturation of the dynamo itself and representation
of the surface field through a mixture of radial and dipole components is likely an inadequate 
description.  Such oversimplification might also confuse the division between wind AML and
internal angular momentum redistribution that is thought significant at ages younger than that of 
the Hyades \citep{sills00}.  Disentangling these effects is likely to require both sophisticated dynamo and
stellar structure models, in addition to coronal MHD wind models such as those presented
here.   However, much empirical progress could still be made through application of the latter
models to stellar surface magnetograms for stars with different rotation periods in order to probe the 
non-orthogonal interplay between magnetic field configuration and AML.

\subsection{Implications for the Short Term Stellar Evolution and Stellar Dynamos}

One can ask whether there might be a feedback mechanism between the rotation, AML and the magnetic 
field structure. It has been shown that re-distribution of stellar magnetic fields affects the spin 
down of stars and reduces their necessity for dynamo saturation \citep{solanki97,holzwarthjardine05}. 
Our results take a step forward in this approach. We suggest that variations in the 
the spot latitudesre-distribution of the surface magnetic field can lead to variations in the wind 
distribution (as observed in the solar case) and as a result, the AML is modified. If the variations 
of the AML affect the motions generating the new magnetic flux (i.e. the convection zone) through small variations 
of the rotation rate, then we speculate that this might be the connection between the corona and the stellar cycle. 

Nevertheless, the feedback from the coronal magnetic field on the solar dynamo has not been studied in detail yet. 
It is possible to perform angular momentum and mass loss rates calculations based on the same method described in this 
paper, but driven by real solar magnetogram data. The predicted variations in AML and mass-loss sould be compared with 
variations of rotation rates in the convection zone measured by GONG and MDI \citep[e.g]{antia08}.


\section{CONCLUSIONS}
\label{sec:conclusion}

We test the effect of star spot latitudinal location and magnetic field strength 
on stellar AML via the stellar wind. Unlike previous studies, we use a model in which 
the wind distribution is dictated by the magnetic field structure. This key feature is crucial in order 
to get a more realistic picture of the corona, in particular in terms of AML rate. 
Even though we use a solar-like wind distribution in this 
work, it seems likely that the relationship between the solar magnetic field expansion and the solar wind is 
universal and can be applied to other Sun-like stars. 

We find that the change in magnetic field structure significantly affects the 
wind structure and as a result, the loss of angular momentum. In particular, we find that the angular 
momentum loss rate is controlled by the mass flux when spots are located at low latitudes. When 
spots are located at high latitudes however, the AML rate is controlled by the 
density increase as a result of fast wind elimination.

This work is a step further in our understanding of the role of magnetic fields in stellar evolution, stellar 
dynamos, and stellar coronal structure. A further investigation of these topics 
involving MHD solutions
driven by surface field maps of real stellar systems \citep[e.g.][]{hussain07,gregory08}, as well as further study of 
idealized cases such as the work presented here, could be of great interest. This work also demonstrates how the detailed knowledge of the physics 
of the solar corona can be implemented for other stars as we study the solar-stellar connection.


\acknowledgments

We would like to thank an unknown referee for useful comments and suggestions. This work has been supported 
by NSF-SHINE ATM-0823592 grant, NASA-LWSTRT Grant NNG05GM44G. Simulation results were obtained using the 
Space Weather Modeling Framework, developed by the Center for Space Environment Modeling, at the University 
of Michigan with funding support from NASA ESS, NASA ESTO-CT, NSF KDI, and DoD MURI.  JJD and VLK were 
funded by NASA contract NAS8-39073 to the {\it Chandra X-ray
Center} (CXC) during the course of this research.


\clearpage


\begin{table}[h]
\begin{tabular}{cccccc}
Co-Latitude & No Spots & 10 & 30 & 60 & 75 \\
\hline
$1kG$ & 5.0 & 3.8 & 4.4 & 5.99 & 7.035  \\
$200G$ & 5.0 & 4.76 & 4.89 & 5.22 & 5.4 \\
\hline
\end{tabular}
\caption{The $g_1^0$ (dipole) component of the magnetic field harmonic expansion for different spot co-latitudes.}
\label{table:1}
\end{table}

\clearpage

\clearpage 

\begin{figure*}[h!]
\centering
\includegraphics[width=3in]{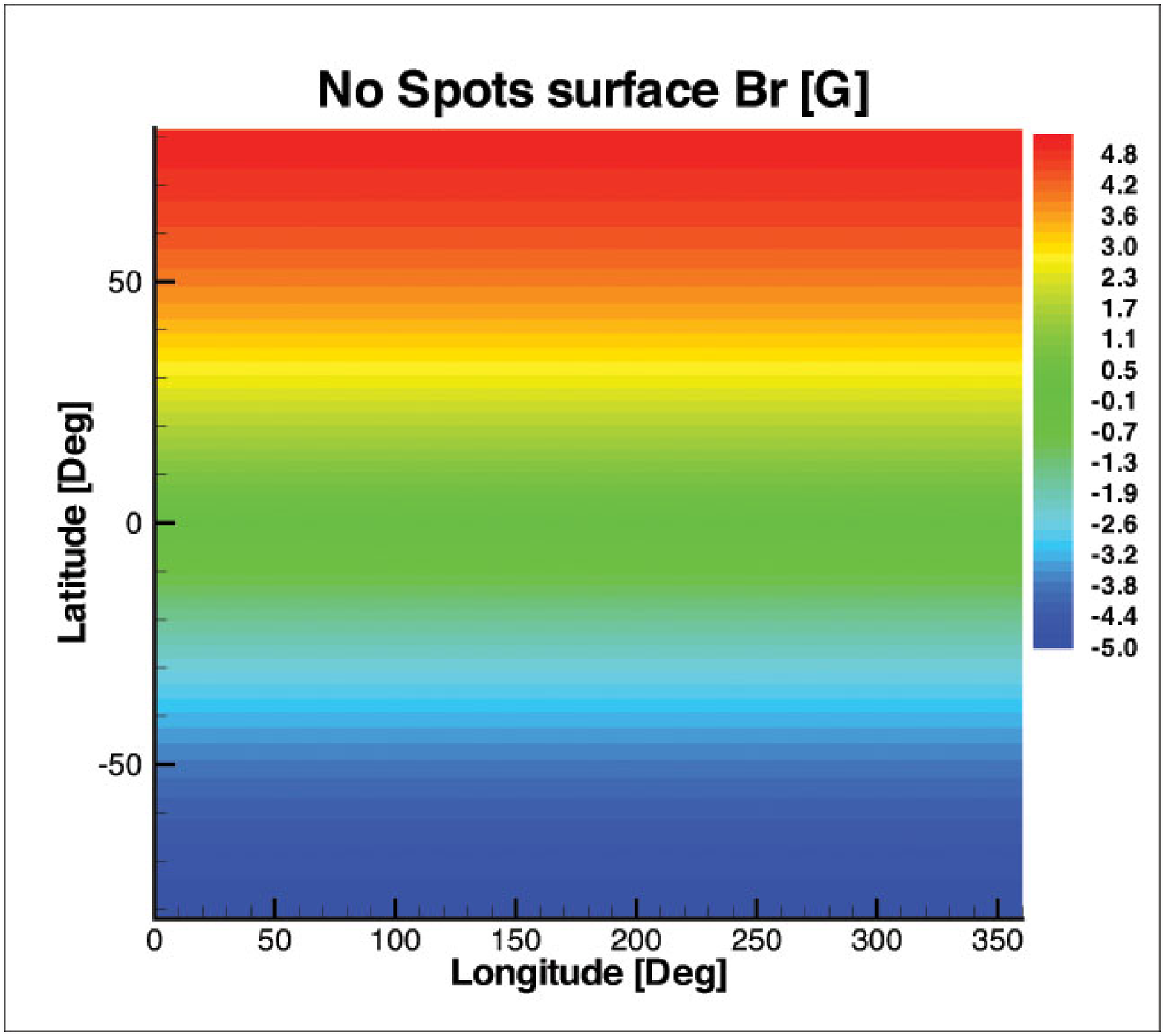}
\includegraphics[width=3in]{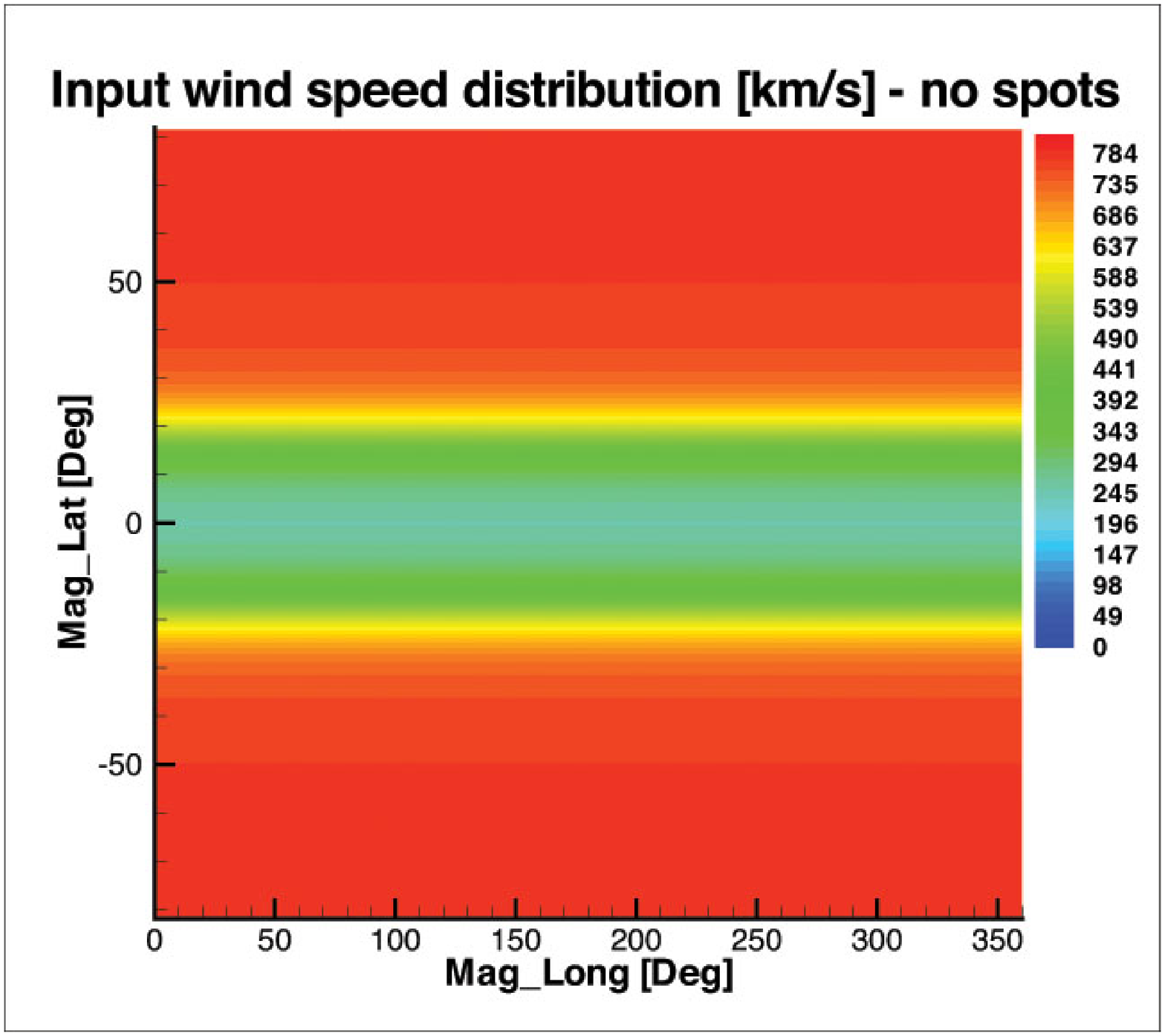} \\
\includegraphics[width=3in]{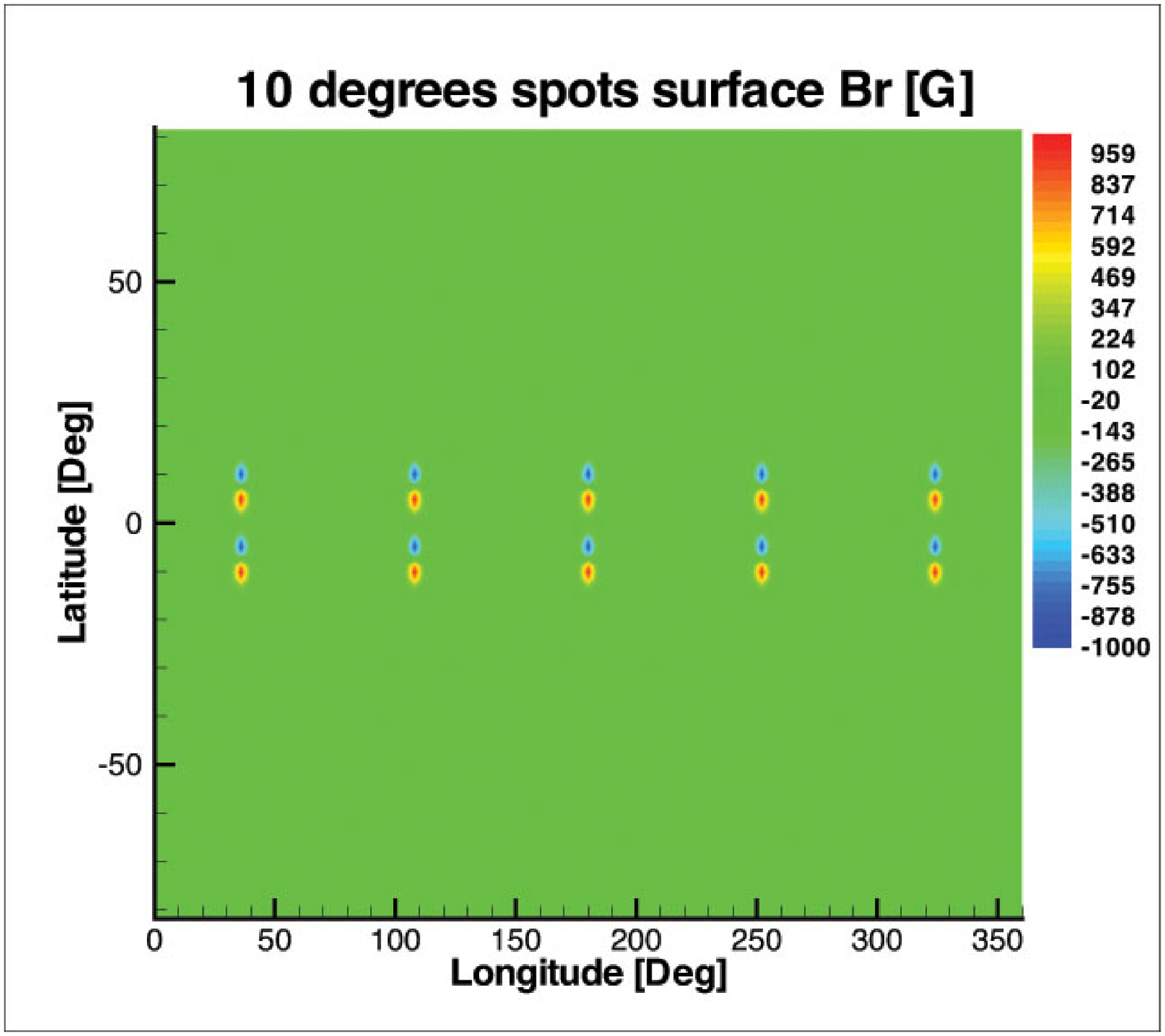}
\includegraphics[width=3in]{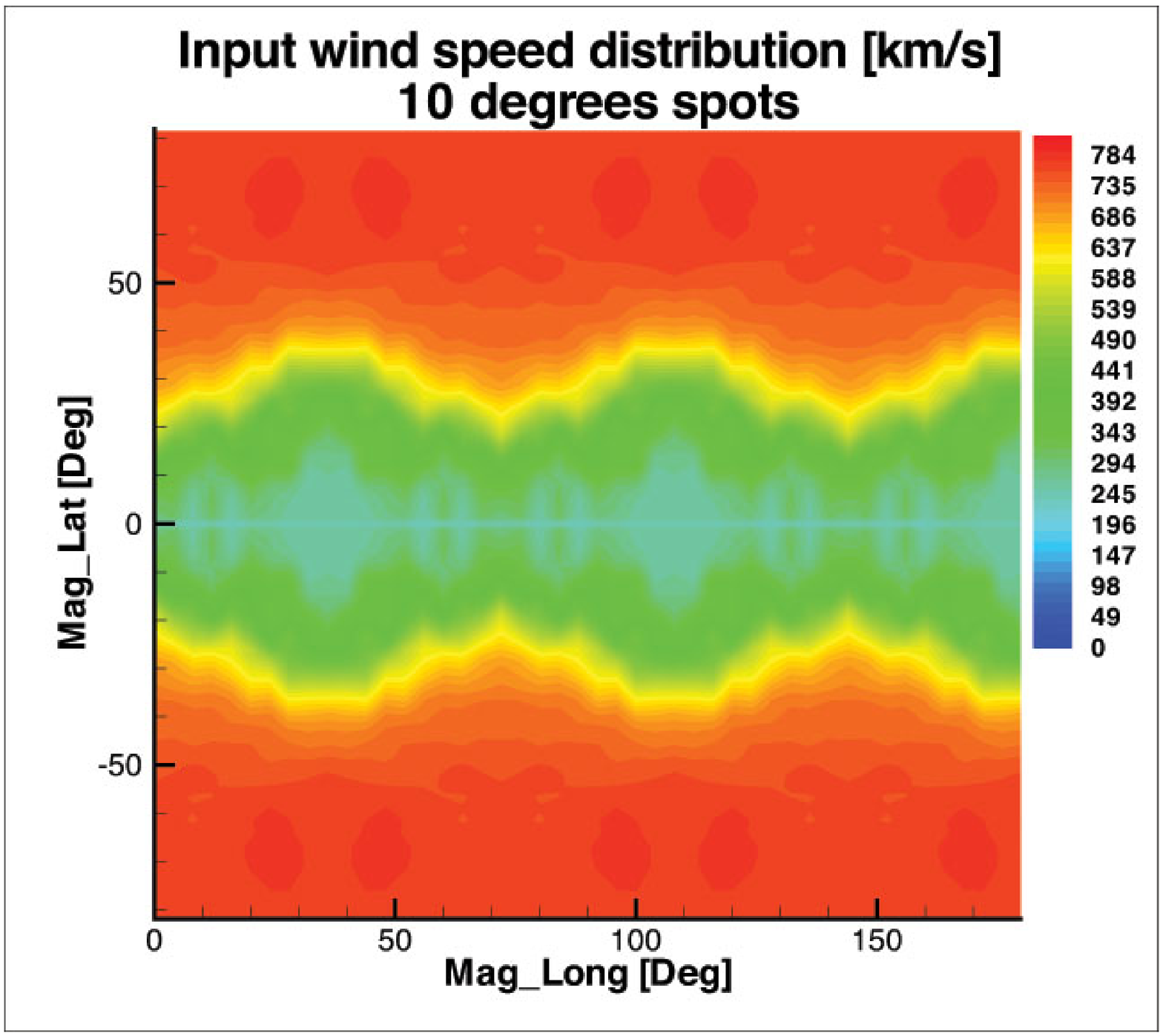} \\
\includegraphics[width=3in]{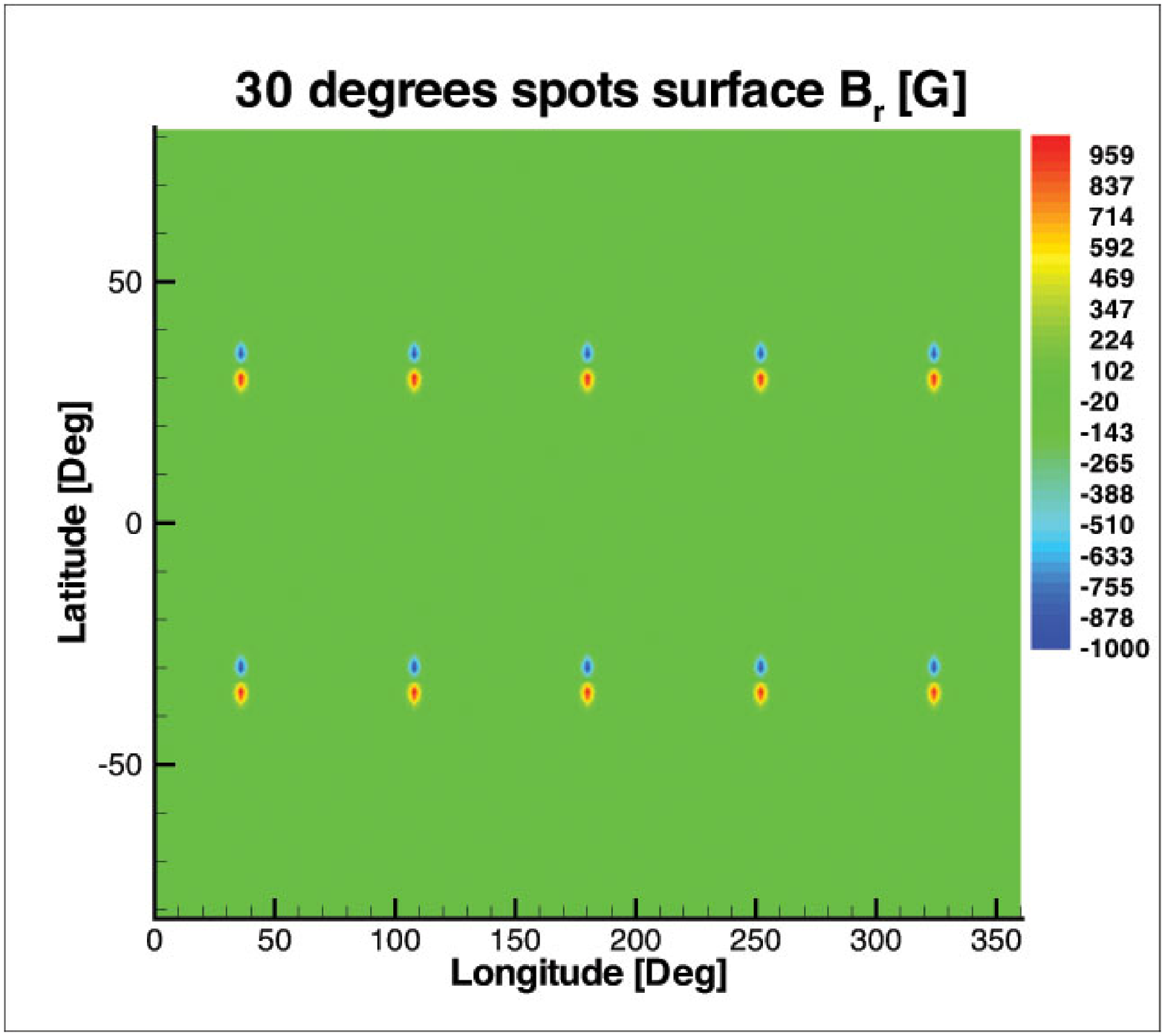}
\includegraphics[width=3in]{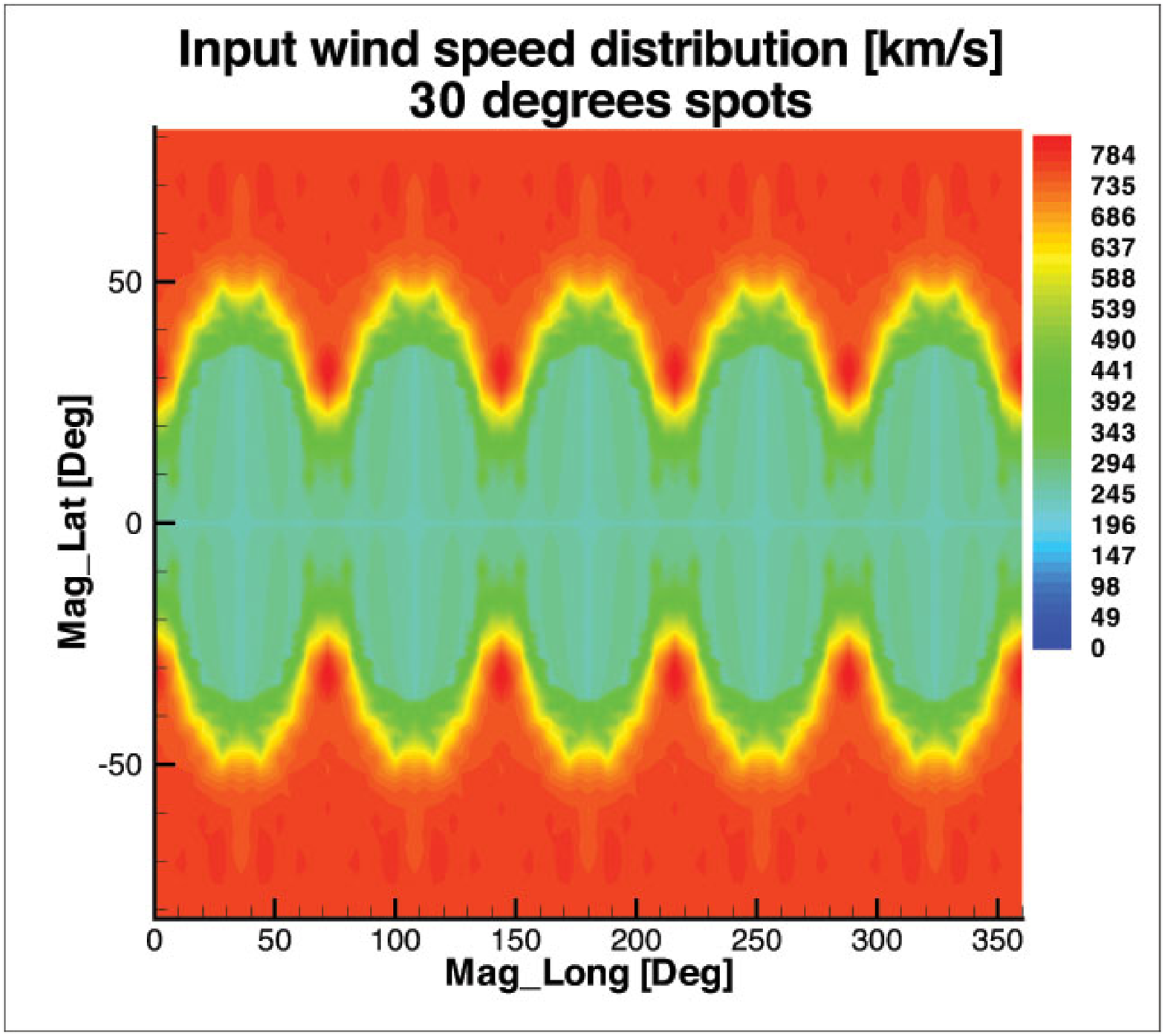}
\end{figure*}
\clearpage 

\begin{figure*}[h!]
\centering
\includegraphics[width=3in]{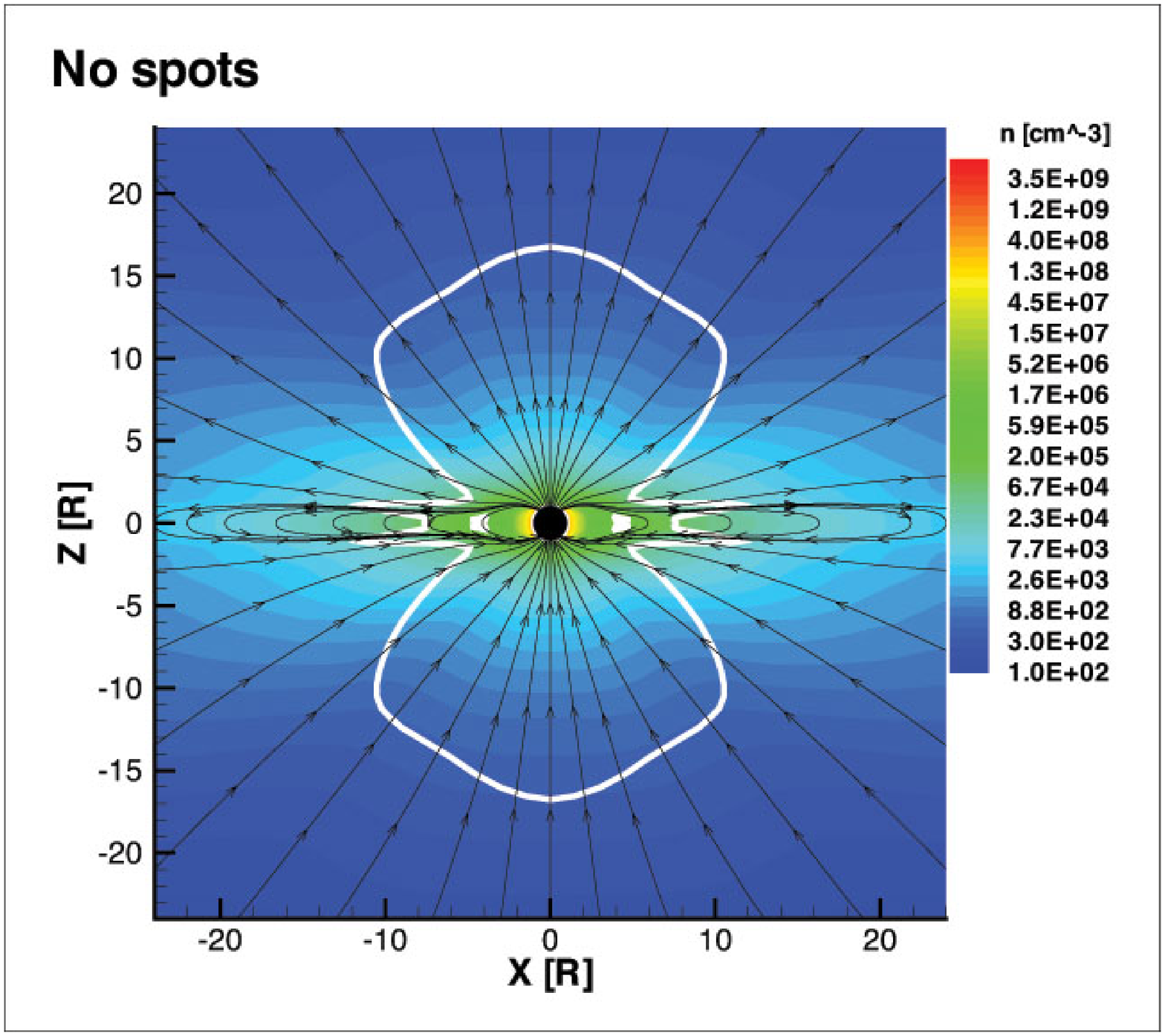}
\includegraphics[width=3in]{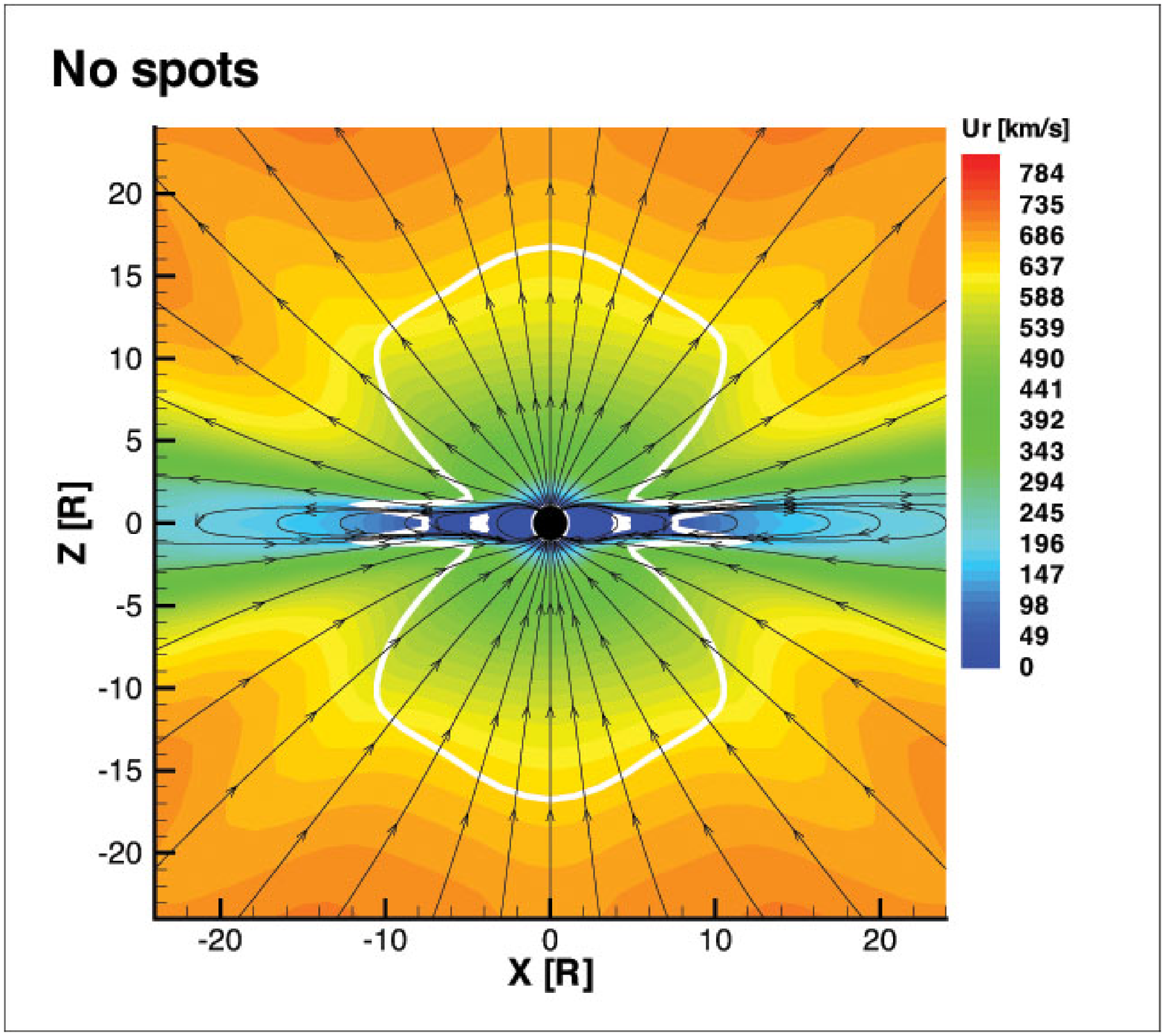} \\
\includegraphics[width=3in]{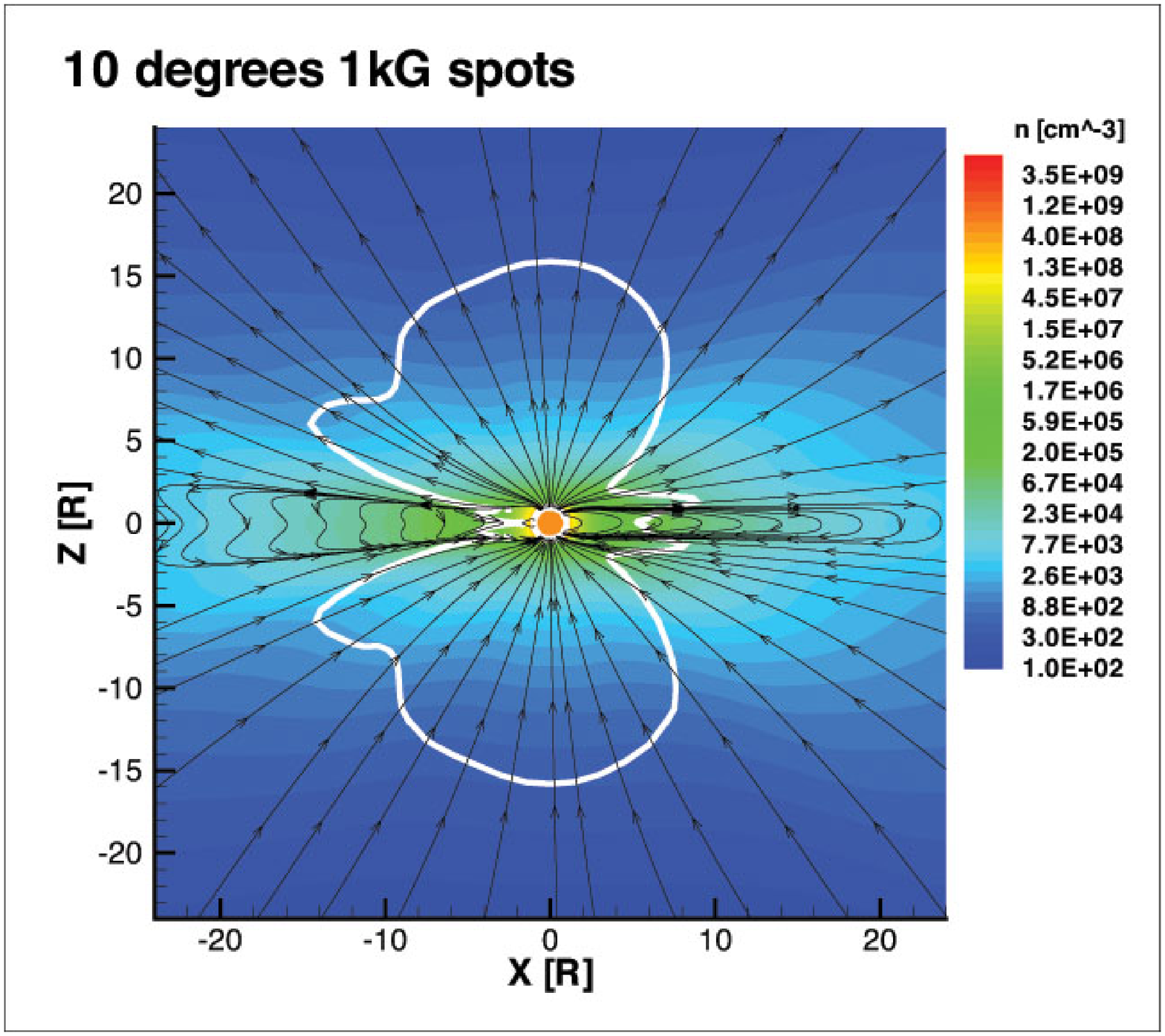}
\includegraphics[width=3in]{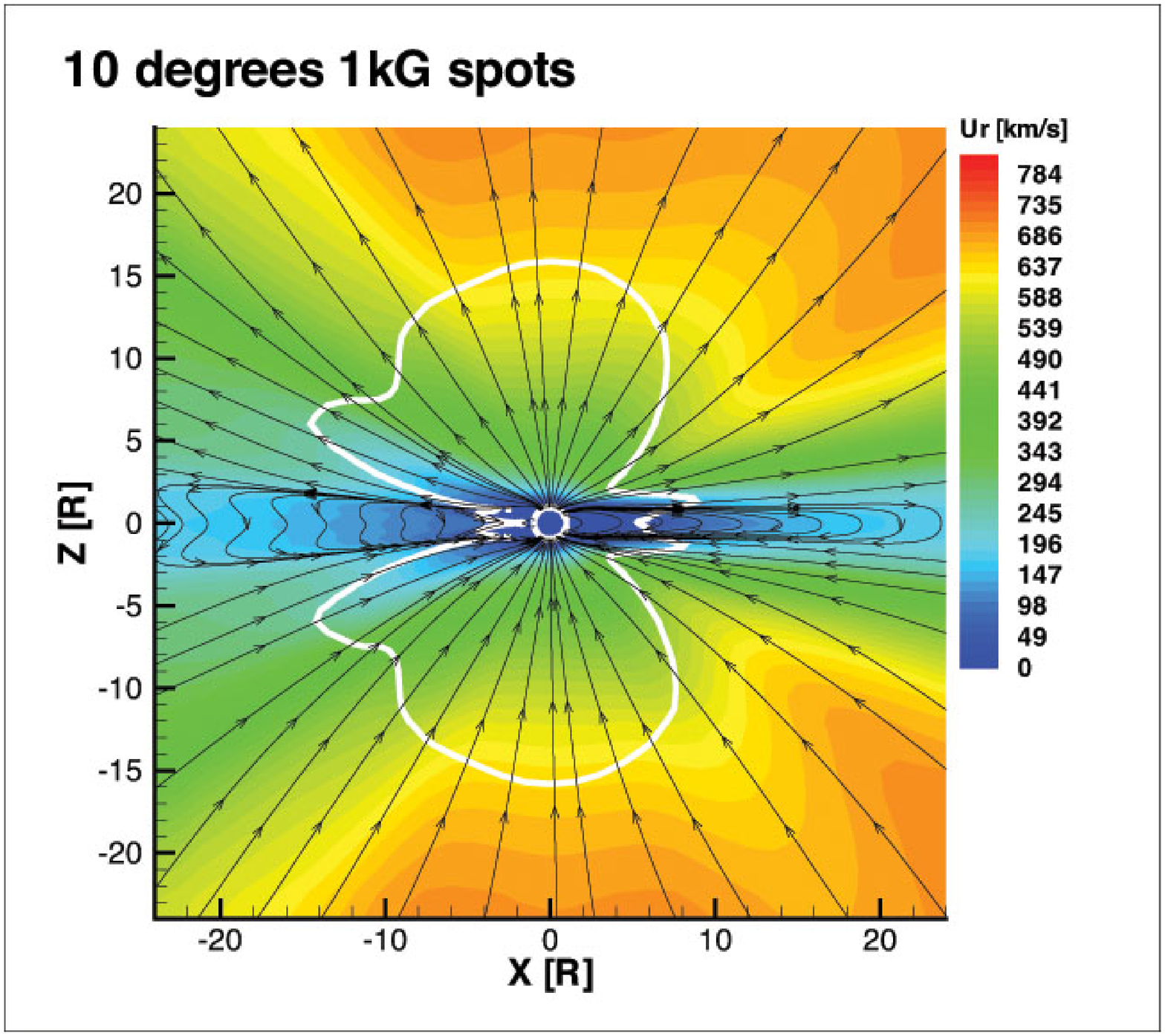} \\
\includegraphics[width=3in]{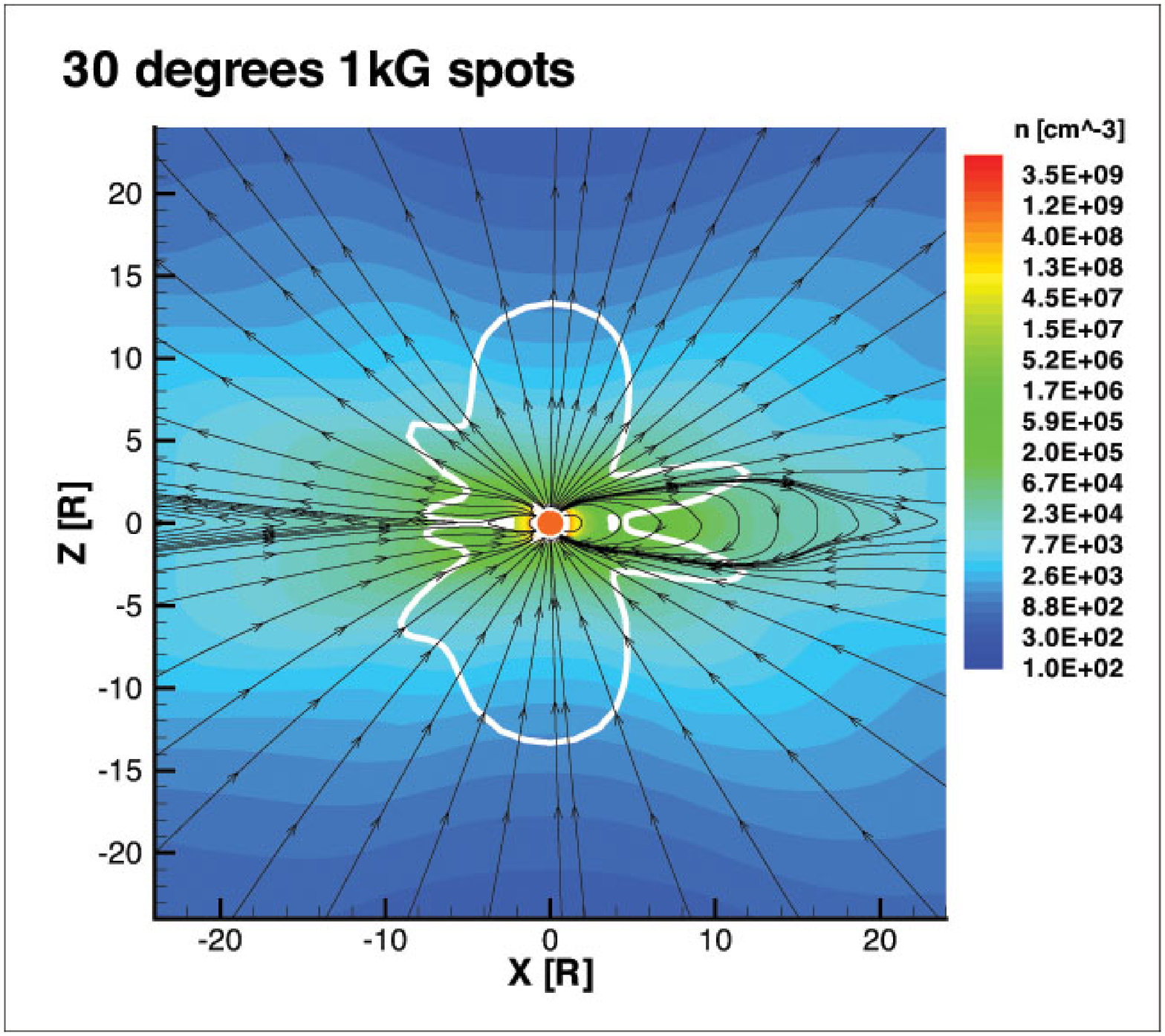}
\includegraphics[width=3in]{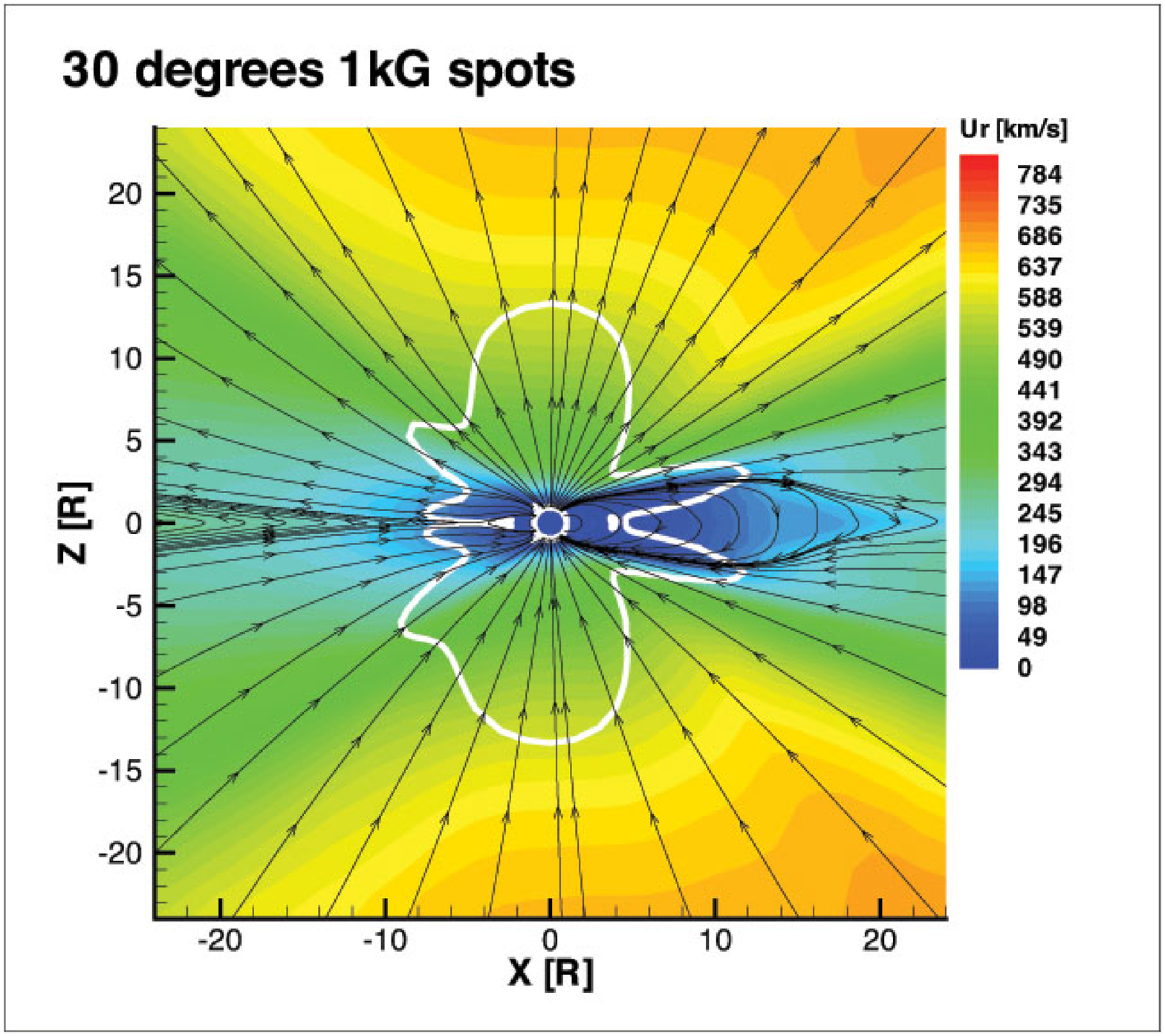}
\end{figure*}
\clearpage 

\end{document}